\documentclass[aps,prd,onecolumn,groupedaddress,showpacs,nofootinbib,amssymb
]{revtex4}
\usepackage[dvips]{graphicx}
\usepackage{amssymb}
\usepackage{amsmath}
\usepackage{epstopdf}
\usepackage{graphicx,,color}
\usepackage{amsfonts}
\usepackage{bm}
\usepackage{cancel}
\usepackage{comment}

\allowdisplaybreaks[4]

\begin{document}

\tolerance=5000



\title{Wormhole models in $f(\textit{R}, \textit{T})$ gravity}

\author{
	Emilio Elizalde$^{1,2}$\thanks{E-mail: elizalde@ieec.uab.es}, 
	Martiros Khurshudyan$^{2,3,4,5}$\thanks{Email: khurshudyan@yandex.ru, khurshudyan@tusur.ru}}

\affiliation{
	$^1$ Consejo Superior de Investigaciones Cient\'{\i}ficas, ICE/CSIC-IEEC,
	Campus UAB, Carrer de Can Magrans s/n, 08193 Bellaterra (Barcelona) Spain \\
	$^{2}$ International Laboratory for Theoretical Cosmology, Tomsk State University of Control Systems 
	and Radioelectronics (TUSUR), 634050 Tomsk, Russia \\
	$^{3}$ Research Division,Tomsk State Pedagogical University, 634061 Tomsk, Russia \\
	$^{4}$ CAS Key Laboratory for Research in Galaxies and Cosmology, Department of Astronomy, University of Science and Technology of China, Hefei 230026, China \\
	$^{5}$ School of Astronomy and Space Science, University of Science and Technology of China, Hefei 230026, China \\
}

\begin{abstract}
	Models of static wormholes within the $f(\textit{R}, \textit{T})$ extended theory of gravity are investigated and, in particular, the family $f(\textit{R}, \textit{T}) = R + \lambda T$, with $T = \rho + P_{r} + 2P_{l}$ being the trace of the energy-momentum tensor. Models corresponding to different relations for the pressure components (radial and lateral), and several equations of state (EoS), reflecting different matter content, are worked out explicitly. The solutions obtained for the shape functions of the generated wormholes obey the necessary metric conditions, as manifested in other studies in the literature. The respective energy conditions reveal the physical nature of the wormhole models thus constructed. It is found, in particular,  that for each of those considered, the parameter space can be divided into different regions, in which the exact wormhole solutions fulfill the null~(NEC) and the weak energy conditions~(WEC), respectively, in terms of the lateral pressure. Moreover, the dominant energy condition~(DEC) in terms of both pressures is also valid, while $\rho + P_{r} + 2P_{l} = 0$. A similar solution for the theory $P_{r} = \omega_{1} \rho + \omega_{2} \rho^{2}$ is found numerically, where $\omega_{1}$ and $\omega_{2}$ are either constant or functions of $r$, leading to the result that the NEC in terms of the radial pressure is also valid. For non-constant $\omega_{i}$ models, attention is focused on the behavior $\omega_{i} \propto r^{m}$. To finish, the question is addressed, how $f(R) = R+\alpha R^{2}$ will affect the wormhole solutions corresponding to fluids of the form $P_{r} = \omega_{1} \rho + \omega_{2} \rho^{2}$, in the three cases mentioned above. Issues concerning the nonconservation of the matter energy-momentum tensor, the stability of the solutions obtained, and the observational possibilities for testing these models are discussed in the last section.
\end{abstract}

\pacs {}

\maketitle

\section{Introduction}\label{sec:INT}

It should come as no surprise that, in the recent literature, a large number of works are devoted to modified theories of gravity, for these theories have proven to be very efficient at solving some important problems that General Relativity has to face up. In particular, the accelerated expansion of the Universe, which, when General Relativity is considered, has to rely necessarily on a dark energy component of unknown nature. This was one of the first motivations that led to formulate modifications of the General Relativity paradigm. A key aspect of such modifications, when aiming at cosmological applications, is to obtain a convincing model for a repulsive force, which would provide a solution fitting the astronomical observations; i.e, able to explain, to start with, the accelerated expansion of the recent past, present, and future Universe. Modified theories of gravity have been  intensively tested in light of cosmological and astrophysical observational data, coming out of various sources. Moreover, they had been used intensively to construct viable models of the early Universe and of cosmic inflation. On top of that, they have been used, too, to model compact objects and to understand the physics of singularities. Each modification of gravity brings along its own interpretation of the energy content of the Universe, responsible for its dynamics and specific physical properties. For instance, in the case of modifications of the geometrical part of the theory, as it happens with f(R) and f(T) gravity theories~(which eventually can be associated to dark energy), the part of the energy source is going to have a geometrical origin. Some discussion of these topics can be found in Refs.~\cite{Nojiri:2017ncd}~-~\cite{Bamba:2008ut}. On the other hand, extended theories of gravity can be constructed not only in the way already mentioned of modifying the geometry, but also from the consideration of extra material contributions. We could expect that these material corrections would come from the existence of imperfect fluids. On the other hand, quantum effects, such as particle production can also be a motivation to consider matter content modified theories of gravity. Some relevant studies on this topic can be found in Refs.~\cite{Harko:2011}~-~\cite{Baffou:2015}.

In general relativity, a wormhole is a hypothetical object that is able to connect asymptotic regions of a single Universe. This imaginary object can work as a tunnel, which may even connect two distinct Universes. The concept of a wormhole is a most popular and intensively studied one in General Relativity research, and in modified theories of gravity, too. The minimal surface area of the attachment connecting the two regions is known as the throat of the wormhole. In spite of the intensive studies, its nature is not  completely understood, yet. In the literature, wormholes are usually classified into two separate groups, namely of static and of dynamic wormholes, respectively. Pioneering work on static wormholes is due to Morris and Thorne, who demonstrated that matter inside them has negative energy, thus violating the null energy condition (NEC)~\cite{Morris:1988}. Trying to develop exact wormhole models, with the possibility either to minimize or even to completely cure this violation, has been one of the most active objects of study in the field, in the last few years.  

Such aim has been fulfilled, in particular, in modified gravity theories~\cite{Jawad:2016}~-~\cite{Zubair:2015}~(to mention a few). One may even expect that primordial wormholes might be present at the very early Universe, where quantum effects play an essential role~\cite{Nojiri:1999}~(and references therein for some initial discussion and results on this issue). On the other hand, if we introduce a scale factor into the original Morris-Thorne metric~(Eq.~(\ref{eq:WHMetric})), this  gives rise to an evolving relativistic wormhole model~(dynamical wormhole model, see for instance~\cite{Bhattacharya:2017} and reference therein). In the recent literature there is an intensive activity related to objects of this kind, as speculating on their existence, in special, with matter satisfying the weak energy condition~(WEC) or the dominant energy condition~(DEC). Moreover, recently too, dynamical wormholes have been treated in terms of a two fluid system. Other interesting studies of traversable wormholes have been carried out in the context of non-equilibrium thermodynamics in the presence of adiabatic particle creation~\cite{Harada:2008}~-~\cite{Pan:2015}~(to mention a few works, only).

Our specific goal in this paper is to construct exact wormhole models assuming different possibilities for their matter content. Concerning the results obtained, we will be able to construct two exact wormhole models using specific connections between radial and lateral pressures. Moreover, we will deduce that the model parameter space can be divided into different regions, where only the  NEC in terms of the radial pressure will be violated~(actually the DEC in terms of the lateral pressure is violated in one of the models, too), while the other energy conditions will remain valid. On the other hand, we will numerically construct, in addition, three new wormhole models, in which all energy conditions will be fulfilled. Only for some values of the model parameters will the DEC in terms of radial and lateral pressures be violated. On the other hand, when considering $f(R) = R + \alpha R^{2}$, instead of $f(R) = R$, then we will find three models with $P_{r} = \omega_{1}\rho + \omega_{2}\rho^{2}$~(where $\omega_{1}$ and $\omega_{2}$ can be also $\propto r^{m}$) for which we will get wormhole solutions, too. Moreover, for one of the solutions to be encountered, only the NEC in terms of $P_{r}$ is violated locally, but far from the throat, while the other energy conditions will be violated everywhere, including the throat  of the wormhole. In the other wormhole solution, the energy conditions will all be violated only locally, far from the throat of the wormhole. This behavior has been observed for a shape function of the form $b(r) = r_{0}/r$  and for different values of the parameter $\alpha$, introduced in accordance with the form of $f(R)$. A general comment may be here in order. We should note that all our solutions will be functions of the $\lambda$ parameter to appear in the corresponding $F(R,T)$ theory to be considered, and also on the
parameters defining the fluids, so that they will be generic functions of the new theories ($T$ plays a definite role). Also, It can be seen from the shape functions for each case that we have results, which are different from those of the $F(R)$ case, being the specific cases discussed in correspondence with  particular values of $\lambda$. In addition, the traversability of
the wormholes will strongly depend on those values of $\lambda$, and this cannot be reached by simply adjusting the values of the parameters defining the fluids.

The contents of the paper are organized as follows.  In Sect.~\ref{sec:WMFE} we present the detailed form of the field equations to be solved, indicating the conditions to satisfy. In Sect.~\ref{sec:MEW}, two exact models for wormholes will be presented, followed with a discussion on the validity of the energy conditions, taking into account two specific forms of the connection between the $P_{l}$ and $P_{r}$ pressures. In Sect.~\ref{seq:M2}, three wormhole candidate solutions for the modified gravity considered, are discussed, an outcome of a precise numerical study of the equations. In this case, too, we will see that the parameter space can be divided into different regions, and will obtain a solution which satisfies all the energy conditions. Furthermore, three wormhole solutions for the fluid models, with given $P_{r}$, in the case of $f(R) = R + \alpha R^{2}$, will be  analyzed in Sect.~\ref{sec:R2}, for $b(r) = r_{0}/r$ as the shape function. Finally, the last section contains a closing discussion and conclusions.

\section{Wormhole metric and the field equations}\label{sec:WMFE}

We will concentrate on one class of $f(\textit{R}, \textit{T})$ theories, namely of the kind given by a total action of the following form~\cite{Harko:2011}
\begin{equation}\label{eq:Action}
S = \frac{1}{16 \pi} \int{ d^{4}x\sqrt{-g} f(\textit{R}, \textit{T}) } + \int{d^{4}x\sqrt{-g} L_{m}},
\end{equation}
where $f(\textit{R}, \textit{T})$ is an arbitrary function of the Ricci scalar, $R$, and of the trace of the energy-momentum tensor, $T$, while $g$ is the metric determinant, and $L_{m}$ the matter Lagrangian density, related to the energy-momentum tensor as
\begin{equation}
T_{ij} = -\frac{2}{\sqrt{-g}} \left[  \frac{\partial (\sqrt{-g} L_{m}) }{\partial g^{ij}} - \frac{\partial}{\partial x^{k}} 
\frac{\partial(\sqrt{-g}L_{m})}{\partial(\partial g^{ij}/\partial x^{k})} \right].
\end{equation}
Now, if we assume that $L_{m}$ depends on the metric components only, we get
\begin{equation}
T_{ij} = g_{ij}L_{m} - 2 \frac{\partial L_{m}}{\partial g^{ij}}.
\end{equation}
On the other hand, variation of the action Eq.~(\ref{eq:Action}) with respect to the metric $g_{ij}$ yields the following field equations
$$f_{R}(\textit{R}, \textit{T}) \left( R_{ij} - \frac{1}{3} R g_{ij} \right) + \frac{1}{6} f(\textit{R}, \textit{T})  g_{ij}  = 8\pi G \left( T_{ij} - \frac{1}{3} T g_{ij} \right) -f_{T}(\textit{R}, \textit{T})  \left( T_{ij} - \frac{1}{3} T g_{ij} \right)$$
\begin{equation}
-f_{T}(\textit{R}, \textit{T})  \left( \theta_{ij} - \frac{1}{3} \theta g_{ij} \right) + \nabla_{i}\nabla_{j} f_{R}(\textit{R}, \textit{T}) ,
\end{equation}
with $f_{R}(\textit{R}, \textit{T}) = \frac{\partial f(\textit{R}, \textit{T}) }{\partial R}$, $f_{T}(\textit{R}, \textit{T}) = \frac{\partial f(\textit{R}, \textit{T}) }{\partial T}$ and
\begin{equation}
\theta_{ij} = g^{ij} \frac{\partial T_{ij}}{\partial g^{ij}}.
\end{equation}

To obtain from here the wormhole solutions, we assume that $L_{m} = - \rho$, in order not to imply the vanishing of the extra force, and $f(\textit{R}, \textit{T}) = R + 2 f(T)$, with $f(T) =  \lambda T$~($\lambda$ is a constant); we rewrite the above equations as follows
\begin{equation}\label{eq:G}
G_{ij} = (8\pi + 2\lambda) T_{ij} + \lambda (2\rho + T)g_{ij},
\end{equation}
$G_{ij}$ being the usual Einstein tensor. A comment is here in order. One can guess that the constraints on $\lambda$ may be important, in some regions at least, from local observations, but these do not exclude the theory as such, as one can realize by looking to the relevant bibliography on the subject, as mentioned above. The fluid approximation is not excluded from the analysis, although its range of applicability is actually relevant under rather extreme conditions. As commonly are, one should not forget, those leading to the existence of wormholes (and of black holes too, by the way, for a different case, but related, at least in this sense). On the other hand, note that on setting $L_m=-\rho$ we are considering a particular case of the theory, which corresponds indeed to well-behaved, specific fluids with admissible equation of state leading to this result. There is no problem with general covariance; this is just an example case, leading to a certain solution. 

In the present section we will briefly address basic issues concerning the metric and the conditions that the solutions of Eq.~(\ref{eq:G}) should satisfy in order to lead to wormholes. In parallel, we will present the final form of the field equations for the class $f(\textit{R}, \textit{T})$ of modified gravities considered. The static spherically symmetric wormhole metric in
Schwarzschild coordinates $(t,r,\theta, \phi)$ is~\cite{Morris:1988}
\begin{equation}\label{eq:WHMetric}
ds^{2} = -U(r) dt^{2} + \frac{dr^{2}}{V} + r^{2}d\Omega^{2},
\end{equation}
where $d\Omega^{2} = d\theta^{2} + sin^{2}\theta d\phi^{2}$ and $V = 1-b(r)/r$. The function $b(r)$ is termed the shape function, since it actually corresponds to the spatial shape of the wormhole. The redshift function $U(r)$ and the shape function $b(r)$ must obey the following conditions~\cite{Morris:1988}:
\begin{enumerate}
\item The radial coordinate $r$ lies between $r_{0} \leq r < \infty$, where $r_{0}$ is the throat radius.
\item At the throat, $r=r_{0}$, $b(r_{0}) = r_{0}$, and for the region outside of the throat, $1- b(r)/r > 0$.
\item $b^{\prime}(r_{0}) < 1$, with $\prime = d/dr$, i.e. should obey the flaring out condition at the throat.
\item For asymptotic flatness of the space-time geometry, the limit $b(r)/r \to  0$, as $|r| \to \infty$ is required.
\item $U(r)$ must be finite and non-vanishing at the throat $r_{0}$.
\end{enumerate}
It is known that if we consider $U(r) =$ const., then we can achieve the de Sitter and the anti-de Sitter asymptotic behaviors. Following Refs.~\cite{Cataldo:2011} and~\cite{Rahaman:2007}, we will consider $U(r) = 1$.  

Now, if we take into account the form of the metric, Eq.~(\ref{eq:WHMetric}), then for the three components of the field equations, Eq.~(\ref{eq:G}), after some algebra we get~\cite{Moraes:2017c}
\begin{equation}\label{eqF1}
\frac{b^{\prime}}{r^{2}} = (8\pi + \lambda)\rho - \lambda (P_{r} + 2P_{l}),
\end{equation}
\begin{equation}\label{eqF2}
-\frac{b}{r^{3}} = \lambda \rho + (8\pi + 3\lambda)P_{r} + 2\lambda P_{l},
\end{equation}
\begin{equation}\label{eqF3}
\frac{b - b^{\prime}r}{2r^{3}} = \lambda \rho + \lambda P_{r} + (8\pi + 4 \lambda) P_{l}.
\end{equation}
To derive the above equations, we have considered an anisotropic fluid displaying a matter content of the form $T^{i}_{j} = \mbox{\rm diag }(-\rho, P_{r},P_{l},P_{l})$, where $\rho = \rho (r)$ is the energy density, while $P_{r}$ and $P_{l}$ are the radial and the lateral pressures, respectively. They are measured orthogonally to the radial direction. The trace $T$ of the energy-momentum tensor turns out to be $T = -\rho + P_{r} + 2P_{l}$. Moreover, Eq.-s~(\ref{eqF1})~-~(\ref{eqF3}) admits the solutions
\begin{equation}\label{eq:rho}
\rho = \frac{b^{\prime} }{r^{2}(8 \pi + 2 \lambda )},
\end{equation}
\begin{equation}\label{eq:Pr}
P_{r} = - \frac{b}{r^{3}(8\pi + 2\lambda )},
\end{equation}
and
\begin{equation}\label{eq:Pl}
P_{l} = \frac{b - b^{\prime}r}{2r^{3}(8\pi + 2\lambda )}.
\end{equation}
The consistency of these equations with the condition on the matter Lagrangian density to depend on the metric components only, and the form for the energy-momentum tensor of the matter field, have been checked. 

In the next section, imposing two relations between $P_{l}$ and $P_{r}$, we will obtain exact wormhole models. On the other hand, in Sect.~\ref{seq:M2} a specific relation between $P_{r}$ and $\rho$ will be assumed and numerical analysis will allow us to construct wormhole models not violating any of the energy conditions. In Ref.~\cite{Moraes:2017c}, for the same   $f(\textit{R}, \textit{T})$ gravity as here, exact wormhole models have been discussed, for $P_{l} = nP_{r}$ and $P_{r} + \omega(r)\rho =  0$. In the last part of this paper, we will deal with wormhole solutions, again obtained numerically, for $f(R) = R + \alpha R^{2}$ and for some fixed relations between $P_{r}$ and $\rho$.

\section{Models of exact wormholes}\label{sec:MEW}
In this section, we  investigate the existence of wormholes and its construction, achieved after involving  different assumptions about the matter content of the cosmological model considered. In particular, we will discuss exact wormhole models derived from two specific assumptions concerning the relation between $P_{l}$ and $P_{r}$.

\subsection{Model with $P_{l} = n P_{r} +\alpha  P_{r}^2$ }

As first model, we assume that the pressures $P_{l}$ and $P_{r}$ are related as
\begin{equation}\label{eq:Pl1}
P_{l} = n P_{r} + \alpha P_{r}^{2},
\end{equation}
being $n$ and $\alpha$ two constants. This relation can be understood as a part of a more general one, namely $P_{l} = \sum{n_{i}P_{r}^{i}}$, a polynomial presentation of the dependence between the pressures. In this sense, we simply restrict our attention to the model with only the first two terms of this polynomial. The form of $P_{l}$ given Eq.~(\ref{eq:Pl1}), together with Eq.~(\ref{eq:Pr}) and Eq.~(\ref{eq:Pl}), allows us to determine the form of $b(r)$, which in this model reads 
\begin{equation}\label{eq:brM1}
b(r) = \frac{1}{A r^{-2 n-1} + b_{1}  r^{-3} },
\end{equation}
where $A$ is an integration constant, while $b_{1} = \frac{\alpha }{2 (\lambda +4 \pi ) (n-1)}$. On the other hand, the form of $b(r)$ allows to obtain the final forms of $\rho$, $P_{r}$ and $P_{l}$, as follows
\begin{equation}\label{eq:rhoM1}
\rho = \frac{A (2 n+1) r^{2 n+2}+3 b_{1} r^{4 n}}{2 (\lambda +4 \pi ) \left(A r^2+ b_{1} r^{2 n}\right)^2},
\end{equation}
\begin{equation}\label{eq:PrM1}
P_{r} = \frac{1}{-2 A (\lambda +4 \pi ) r^{2-2 n}-2 b_{1} (\lambda +4 \pi )},
\end{equation}
and 
\begin{equation}\label{eq:PlM1}
P_{l} = \frac{\alpha -2 A (\lambda +4 \pi ) n r^{2-2 n}-2 b_{2} (\lambda +4 \pi ) n}{4 (\lambda +4 \pi )^2 \left(A r^{2-2 n}+b_{2}\right)^2}.
\end{equation}

\begin{figure}[h!]
 \begin{center}$
 \begin{array}{cccc}
\includegraphics[width=80 mm]{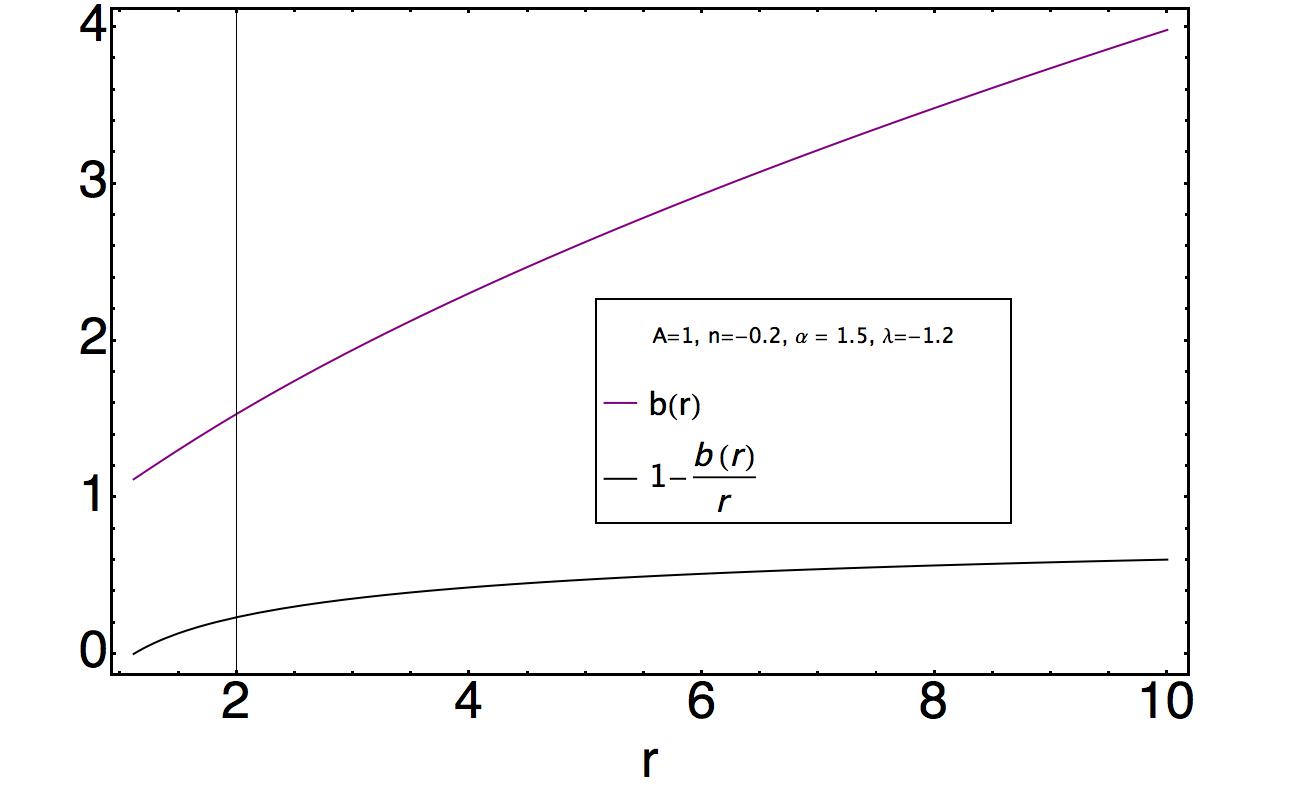} &&
\includegraphics[width=75 mm]{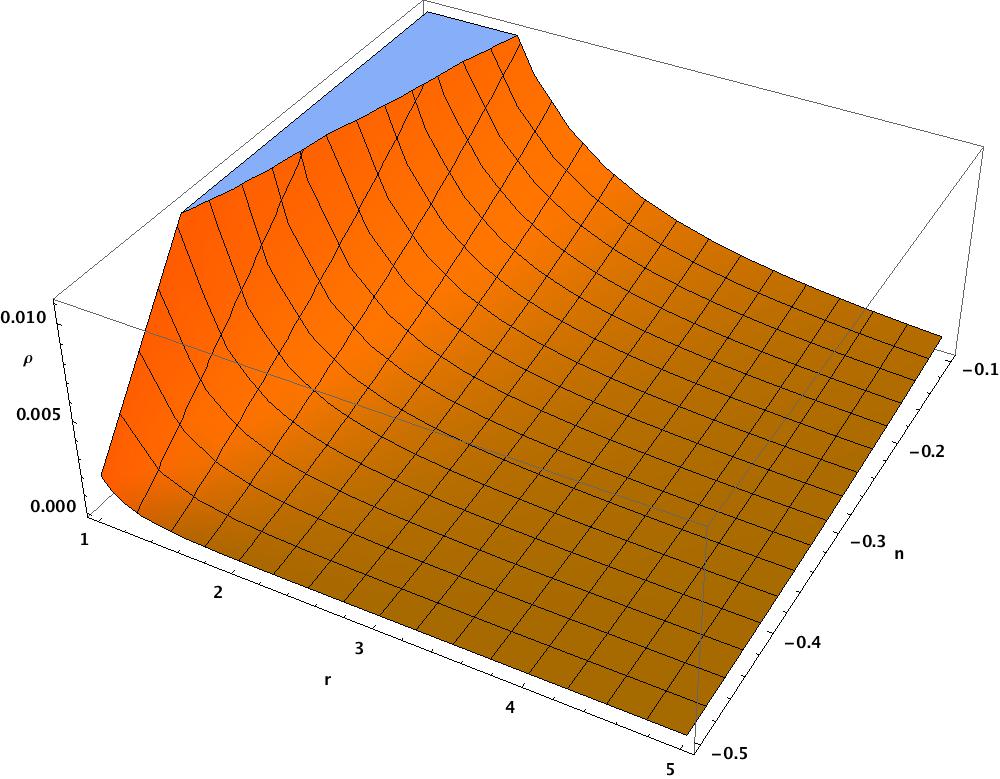}
 \end{array}$
 \end{center}
\caption{The  behavior of the shape function $b(r)$ for model $1$ is depicted on the lhs plot. The same plot clearly shows that the solution, Eq.~(\ref{eq:brM1}), for $b(r)$, satisfies $1- b(r)/r > 0$, when $r > r_{0}$. From the rhs plot we see that $\rho \geq 0$, for $A = 1.0,$ $\alpha = -0.5$, $ \lambda = -1.2$, and for different values of the parameter $n$ ($r$ is given in $[km]$).}
 \label{fig:Fig1}
\end{figure}

The left plot of Fig.~(\ref{fig:Fig1}) describes the graphical behavior of the shape function $b(r)$, which clearly satisfies the well-known condition  $b(r) < r$, as it should be. On the other hand, from the same plot it is clear that the solution of $b(r)$ satisfies $1- b(r)/r > 0$, for $r > r_{0}$. In this particular case, the throat of the wormhole is formed at $r_{0} \approx 1.1144$.  The flaring out condition at the throat has also been checked, giving $b^{\prime}(r_{0}) \approx 0.494$. This analysis proves that we have  constructed an exact viable model of a wormhole, described by  Eqs.~(\ref{eq:brM1})~-~(\ref{eq:PlM1}). Moreover, the rhs plot of Fig.~\ref{fig:Fig1} visually shows that $\rho \geq 0$, for $A = 1.0,$ $\alpha = -0.5$, $ \lambda = -1.2$, and for different values of the parameter $n$.

Now, let us concentrate our attention on the graphical behavior of the other energy conditions, obtained as follows:
\begin{equation}\label{eq:NECR}
\rho + P_{r} = \frac{A n r^{2 n+2}+ b_{1} r^{4 n}}{(\lambda +4 \pi ) \left(A r^2+ b_{1} r^{2 n}\right)^2},
\end{equation}
\begin{equation}\label{eq:NECL}
\rho + P_{l} = \frac{2 A (\lambda +4 \pi ) (n+1) r^{2 n+2}+r^{4 n} (\alpha -2 b_{1} (\lambda +4 \pi ) (n-3))}{4 (\lambda +4 \pi )^2 \left(A r^2+b_{1} r^{2 n}\right)^2},
\end{equation}
\begin{equation}\label{eq:DECR}
\rho - P_{r} = \frac{A (n+1) r^{2 n+2}+2 b_{1} r^{4 n}}{(\lambda +4 \pi ) \left(A r^2+ b_{1} r^{2 n}\right)^2},
\end{equation}
and 
\begin{equation}\label{eq:DECL}
\rho - P_{l} = \frac{2 A (\lambda +4 \pi ) (3 n+1) r^{2 n+2}+r^{4 n} (2 b_{1} (\lambda +4 \pi ) (n+3)-\alpha )}{4 (\lambda +4 \pi )^2 \left(A r^2+b_{1} r^{2 n}\right)^2}.
\end{equation}

\begin{figure}[h!]
 \begin{center}$
 \begin{array}{cccc}
\includegraphics[width=80 mm]{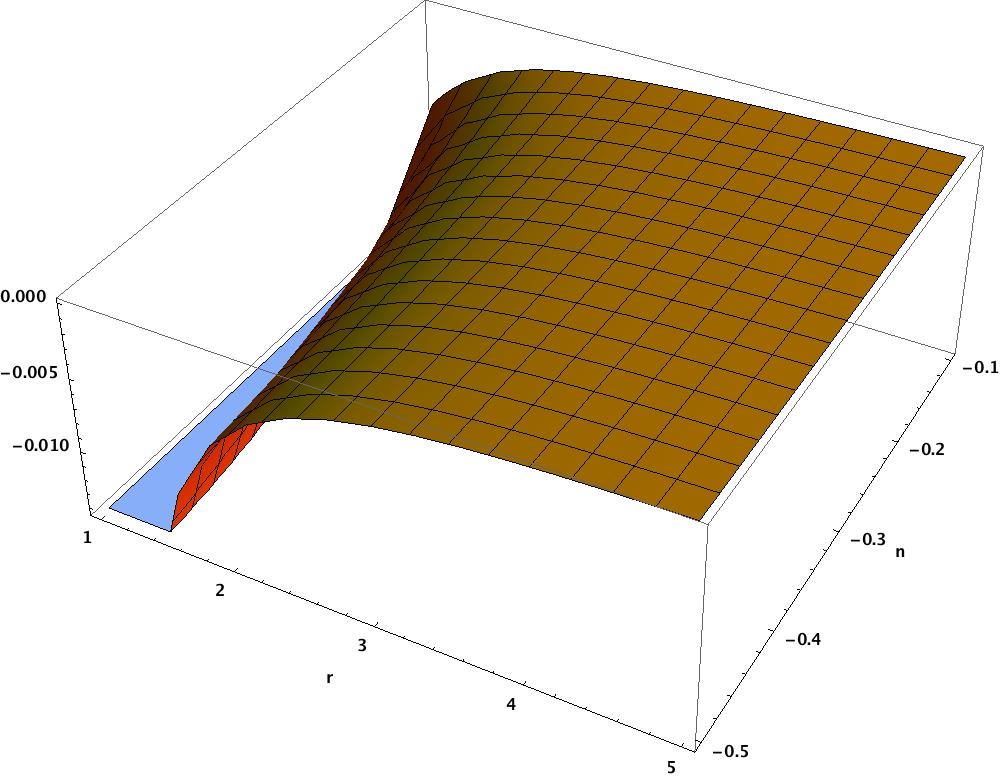} &&
\includegraphics[width=80 mm]{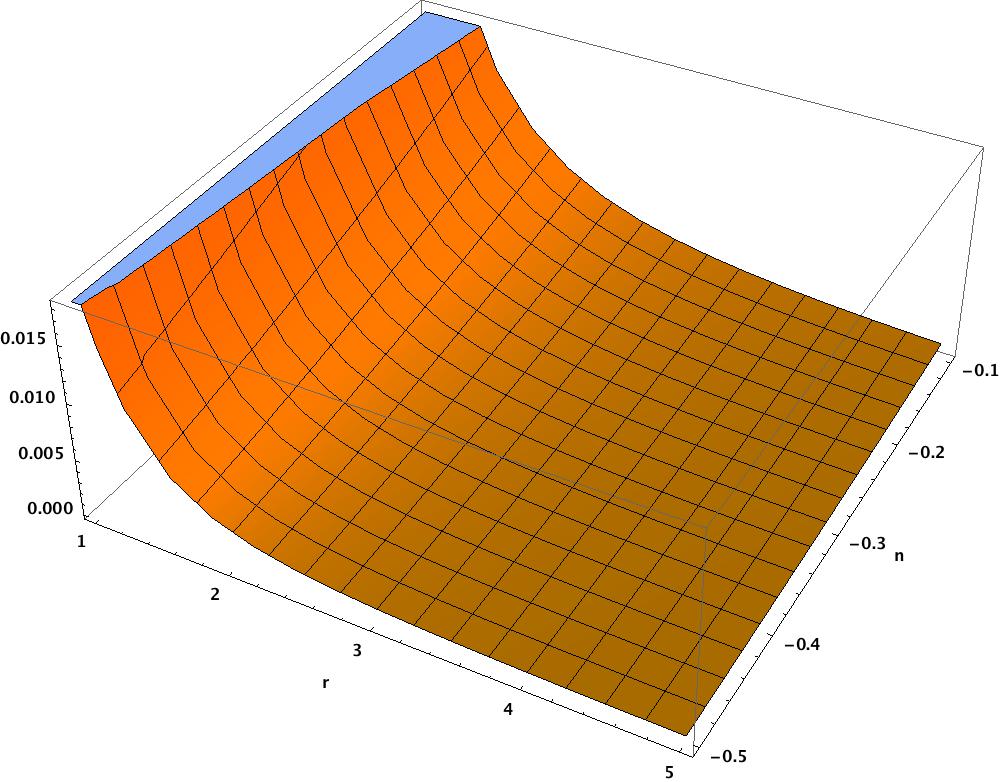} \\
\includegraphics[width=80 mm]{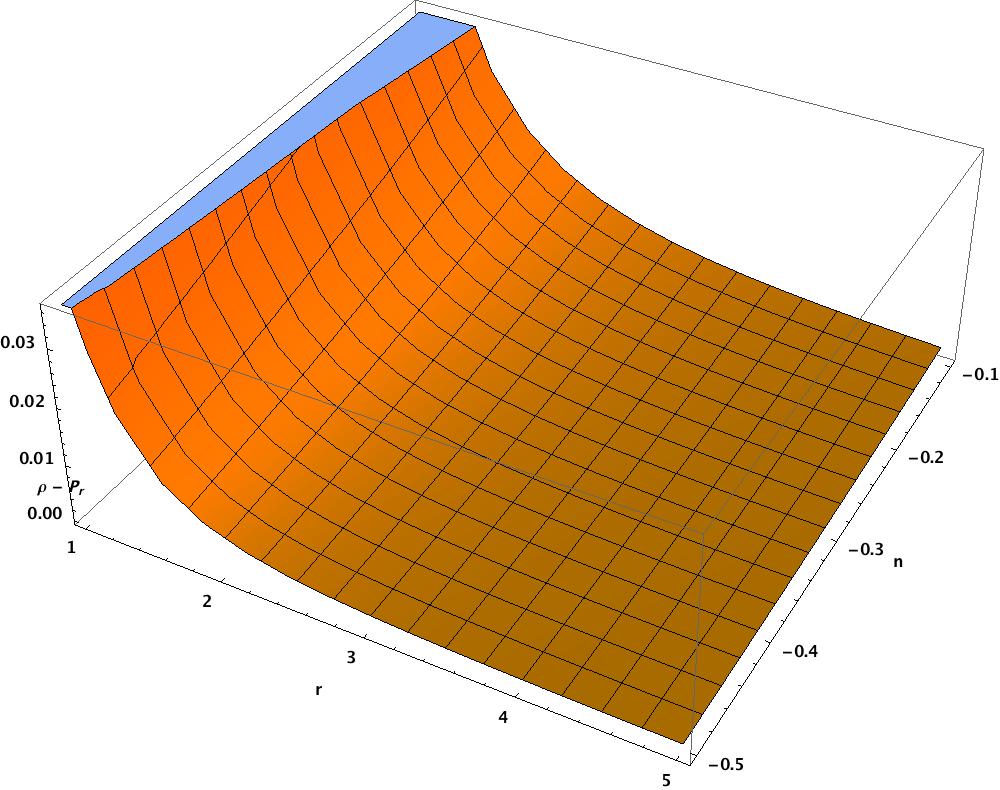} &&
\includegraphics[width=80 mm]{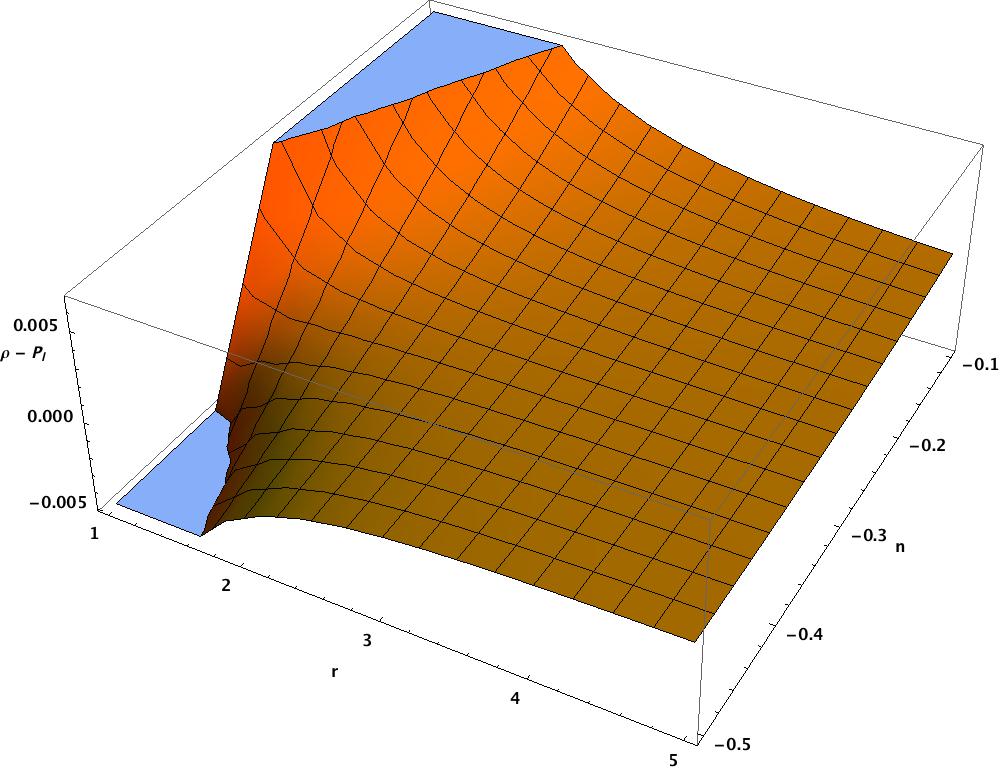}
 \end{array}$
 \end{center}
\caption{The behavior of the null energy condition~(NEC i.e $\rho + P_{r}$) in terms of the $P_{r}$ pressure, as given by Eq.~(\ref{eq:NECR}), for the model $1$ is depicted on the upper-left plot. The behavior of NEC~($\rho + P_{l} $) in terms of the $P_{l}$ pressure, as given by Eq.~(\ref{eq:NECL}), is represented on the upper-right plot. The bottom panel corresponds to the dominant energy condition~(DEC), $\rho - P_{r}$ and $\rho - P_{l}$, in terms of $P_{r}$ and $P_{l}$, as given by Eq.~(\ref{eq:DECR}) and Eq.~(\ref{eq:DECL}), respectively. The bottom-left and the bottom-right plots are for DEC, again in terms of $P_{r}$ and $P_{l}$, respectively. $r$ is in $[km]$, while pressure and the energy densities are dimensionless.}
 \label{fig:Fig2}
\end{figure}

The top panel of Fig.~\ref{fig:Fig2} is a surface plot of thel behavior of the NEC in terms of $P_{r}$~(upper-left) and $P_{l}$~(upper-right), while the DEC in terms of $P_{r}$~(bottom-left) and $P_{l}$~(bottom-right), respectively, is depicted on the bottom panel of Fig.~(\ref{fig:Fig2}). Analysis of the four plots proves that it is possible to choose the model parameters in such a way that we can get a wormhole model for which only the NEC in terms of $P_{r}$ and the DEC in terms of $P_{l}$ are violated, while the WEC in terms of $P_{l}$ remains valid. It is also possible to have that only the NEC in terms of $P_{r}$ is violated, while the other energy conditions are still valid. In other words, we have been able in this case to construct an exact wormhole model violating the NEC only, while maintaining the validity of the WEC.

\subsection{Model with $P_{l} = n P_{r} +  \alpha  r^{m} P_{r} ^2$}

We are also able to find exact wormhole solution by considering the following relation between the  pressures $P_{l}$ and $P_{r}$, namely
\begin{equation}
P_{l} = n P_{r} +  \alpha  r^{m} P_{r} ^2.
\end{equation}
 In this case the shape function reads 
\begin{equation}\label{eq:brM2}
b(r) = \frac{1}{A r^{-2 n-1}+ b_{1} r^{m-3}},
\end{equation}
where $b_{1} = \frac{\alpha }{(\lambda +4 \pi ) (m+2 n-2)}$. 

\begin{figure}[h!]
 \begin{center}$
 \begin{array}{cccc}
\includegraphics[width=80 mm]{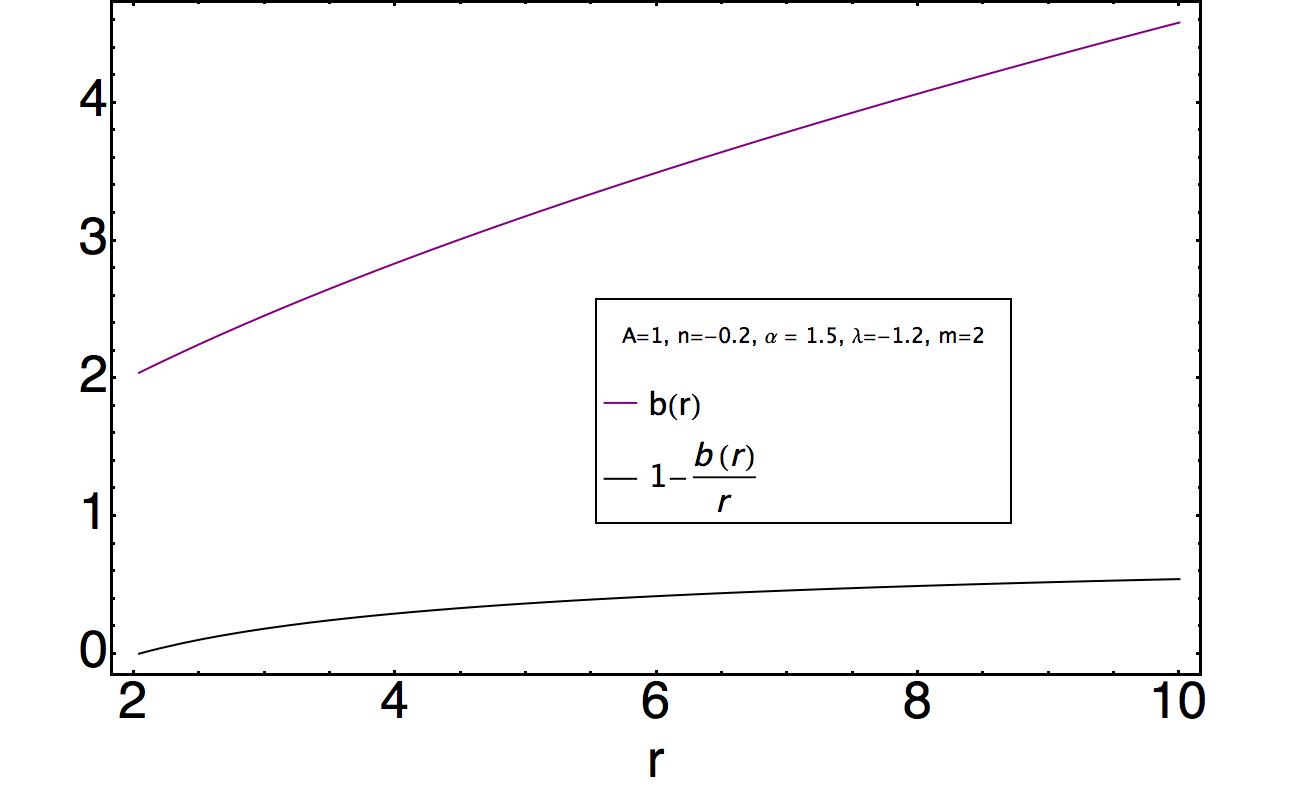} &&
\includegraphics[width=75 mm]{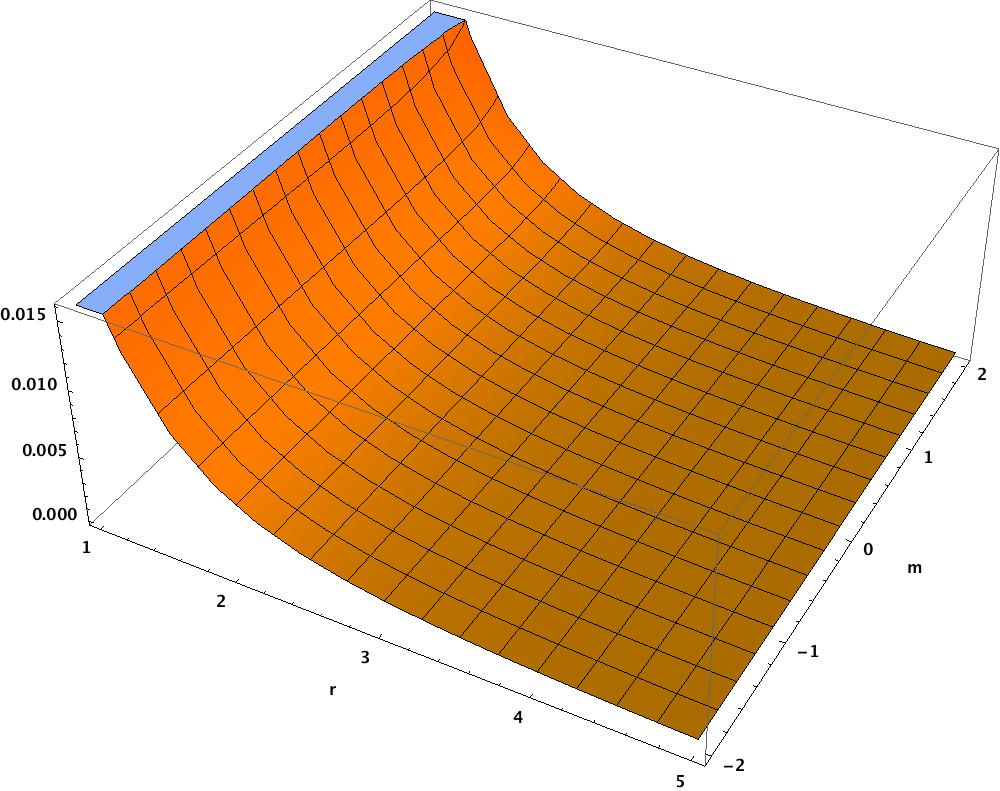}
 \end{array}$
 \end{center}
\caption{The graphical behavior of the shape function $b(r)$ for model $2$ is depicted on the lhs. The same plot demonstrates that the solution, Eq.~(\ref{eq:brM2}), for $b(r)$ satisfies $1- b(r)/r > 0$, for $r > r_{0}$. The rhs plot shows that $\rho \geq 0$ for $A = 1.0,$ $n=-0.2,$ $\alpha = 1.5$, $ \lambda = -1.2$, and for different values of the parameter $m$. $r$ is in $[km]$.}
 \label{fig:Fig3}
\end{figure}

The behavior of $b(r)$ isdepicted on the lhs plot of Fig.~\ref{fig:Fig3}. Similar to the previous model, we see that for appropriate values of the parameters, it is possible to satisfy $b(r) < r$ and $1- b(r)/r > 0$, for $r > r_{0}$. The flaring out condition at the throat has also been checked, giving $b^{\prime}(r_{0}) \approx 0.468$.  The right plot of Fig.~(\ref{fig:Fig3}) shows that $\rho \geq 0$,  for $A = 1, $ $n=-0.2,$ $\alpha = 1.5,$ $\lambda = -1.2$, for different values of the parameter $m$. Using the form of $b(r)$ obtained, Eq.~(\ref{eq:brM2}), we can calculate 
\begin{equation}
\rho = \frac{A (2 n+1) r^{2 n+2}-b_{1} (m-3) r^{m+4 n}}{2 (\lambda +4 \pi ) \left(A r^2+b_{1} r^{m+2 n}\right)^2},
\end{equation}
\begin{equation}
P_{r} = \frac{1}{-2 A (\lambda +4 \pi ) r^{2-2 n}-2 b_{1} (\lambda +4 \pi ) r^m},
\end{equation}
and
\begin{equation}
P_{l} = \frac{r^{m+4 n} (\alpha -2 b_{1} (\lambda +4 \pi ) n)-2 A (\lambda +4 \pi ) n r^{2 n+2}}{4 (\lambda +4 \pi )^2 \left(A r^2+b_{1} r^{m+2 n}\right)^2}.
\end{equation}

\begin{figure}[h!]
 \begin{center}$
 \begin{array}{cccc}
\includegraphics[width=80 mm]{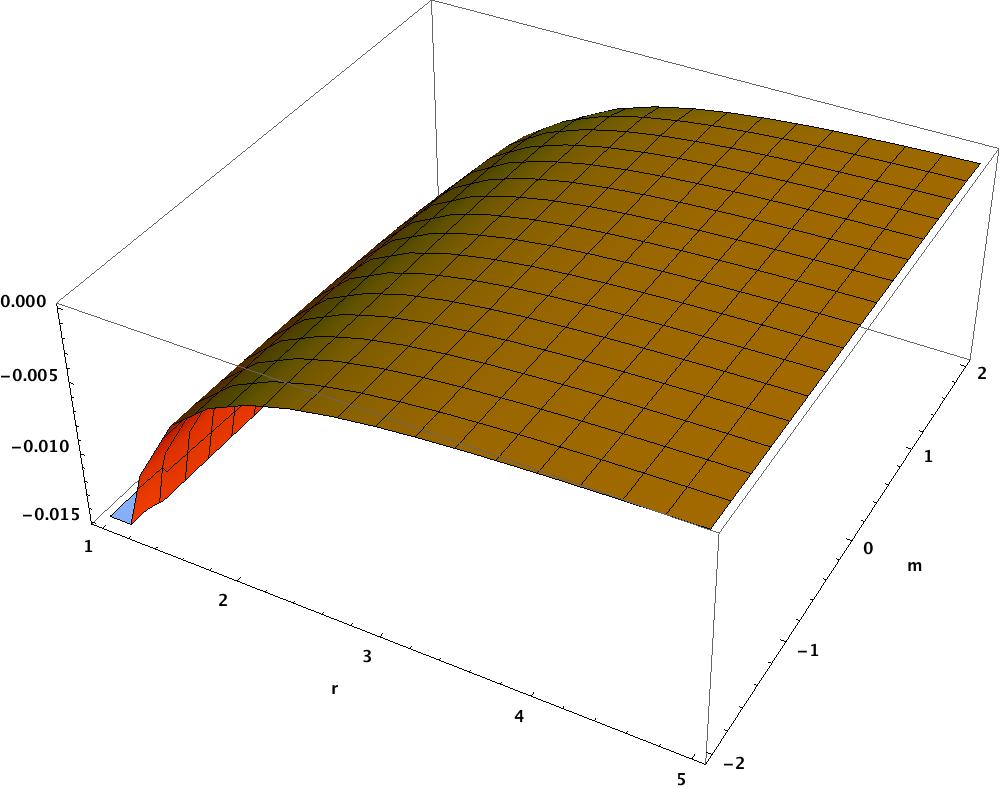} &&
\includegraphics[width=80 mm]{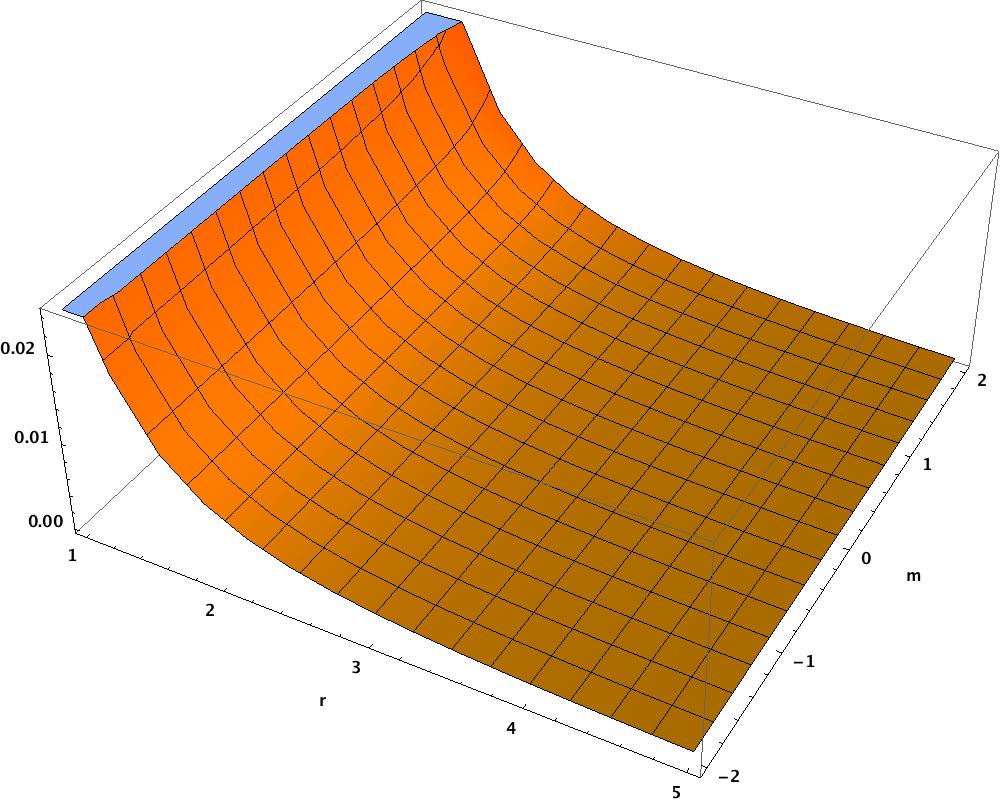} \\
\includegraphics[width=80 mm]{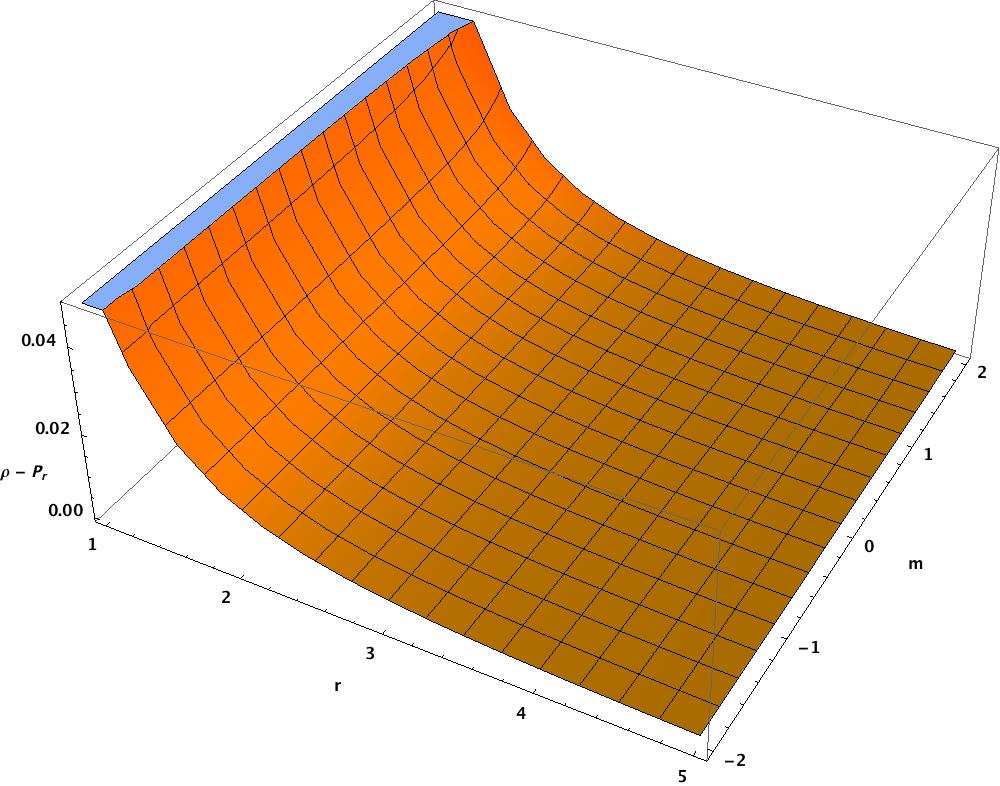} &&
\includegraphics[width=80 mm]{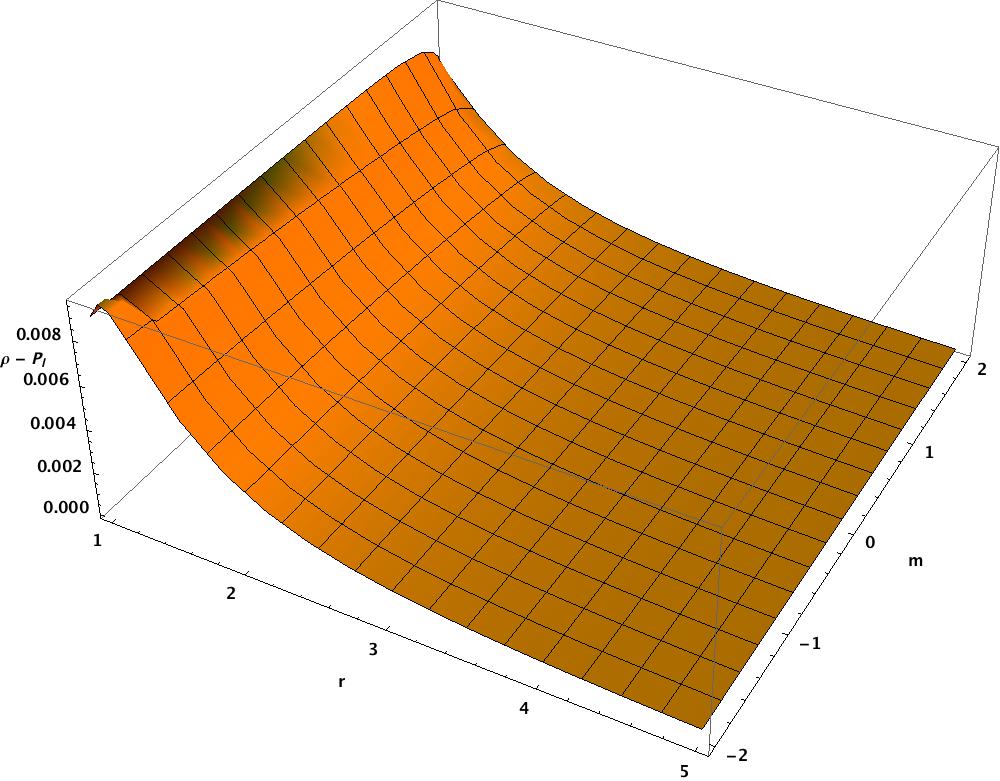}
 \end{array}$
 \end{center}
\caption{The graphical behavior of the null energy condition~(NEC) in terms of the $P_{r}$ pressure, given by Eq.~(\ref{eq:NECR2}), for model $2$ is depicted on the upper-left plot. The graphical behavior of the NEC in terms of the $P_{l}$ pressure, given by Eq.~(\ref{eq:NECL2}), is shown on the upper-right plot. The bottom panel corresponds to the dominant energy condition~(DEC) in terms of the $P_{r}$ and $P_{l}$ pressures, given by Eqs.~(\ref{eq:DECR2}) and~(\ref{eq:DECL2}), respectively. The bottom-left and the bottom-right plots correspond to the DEC in terms of the pressures  $P_{r}$ and $P_{l}$, respectively. $r$ is in $[km]$, while pressure and the energy densities are dimensionless.}
 \label{fig:Fig4}
\end{figure}

In order, to understand what happens now with the energy conditions, let us calculate $\rho + P_{r}$, $\rho + P_{l}$, $\rho - P_{r}$ and $\rho - P_{l}$. We obtain 
\begin{equation}\label{eq:NECR2}
\rho + P_{r} = \frac{2 A n r^{2 n+2}-b_{1} (m-2) r^{m+4 n}}{2 (\lambda +4 \pi ) \left(A r^2+b_{1} r^{m+2 n}\right)^2},
\end{equation}
\begin{equation}\label{eq:NECL2}
\rho + P_{l} = \frac{r^{2 n} \left(2 A (\lambda +4 \pi ) (n+1) r^2+r^{m+2 n} (\alpha -2 b_{1} (\lambda +4 \pi ) (m+n-3))\right)}{4 (\lambda +4 \pi )^2 \left(A r^2+b_{1} r^{m+2 n}\right)^2},
\end{equation}
\begin{equation}\label{eq:DECR2}
\rho - P_{r} = \frac{2 A (n+1) r^{2 n+2}-b_{1} (m-4) r^{m+4 n}}{2 (\lambda +4 \pi ) \left(A r^2+b_{1} r^{m+2 n}\right)^2},
\end{equation}
and
\begin{equation}\label{eq:DECL2}
\rho - P_{l} = \frac{r^{2 n} \left(2 A (\lambda +4 \pi ) (3 n+1) r^2-r^{m+2 n} (\alpha +2 b_{1} (\lambda +4 \pi ) (m-n-3))\right)}{4 (\lambda +4 \pi )^2 \left(A r^2+b_{1} r^{m+2 n}\right)^2}.
\end{equation}

The graphical behavior of the energy conditions given by the equations above  is shown in Fig.~\ref{fig:Fig4}. We see that only the NEC in terms of $P_{r}$~(upper-left plot) is violated. The NEC in terms of $P_{l}$, and the DEC in terms of $P_{l}$ and $P_{r}$ are still valid, yielding a WEC in terms of $P_{l}$ which is valid, too. It should be mentioned that, for both models, $\rho + P_{r}+2P_{l}$ is exactly $0$. Moreover, we have noticed that the parameter $m$ cannot change the nature of the energy condition.

\section{Models with $P_{r} = \omega_{1} \rho + \omega_{2} \rho^{2}$}\label{seq:M2}

In this section we present the results coming from a numerical study of a wormhole model described by the following general expression for the $P_{r}$ pressure
\begin{equation}\label{eq:Fluid0}
P_{r} = \omega_{1} \rho + \omega_{2} \rho^{2}.
\end{equation}
Here $\omega_{1}$ and $\omega_{2}$ can be either constant or are allowed to depend on $r$. Fluids of these types have been intensively, and very successfully, considered in cosmology. 
We start the numerical analysis of the model for the case when $\omega_{1}$ and $\omega_{2}$ are constant.

\begin{figure}[h!]
 \begin{center}$
 \begin{array}{cccc}
\includegraphics[width=80 mm]{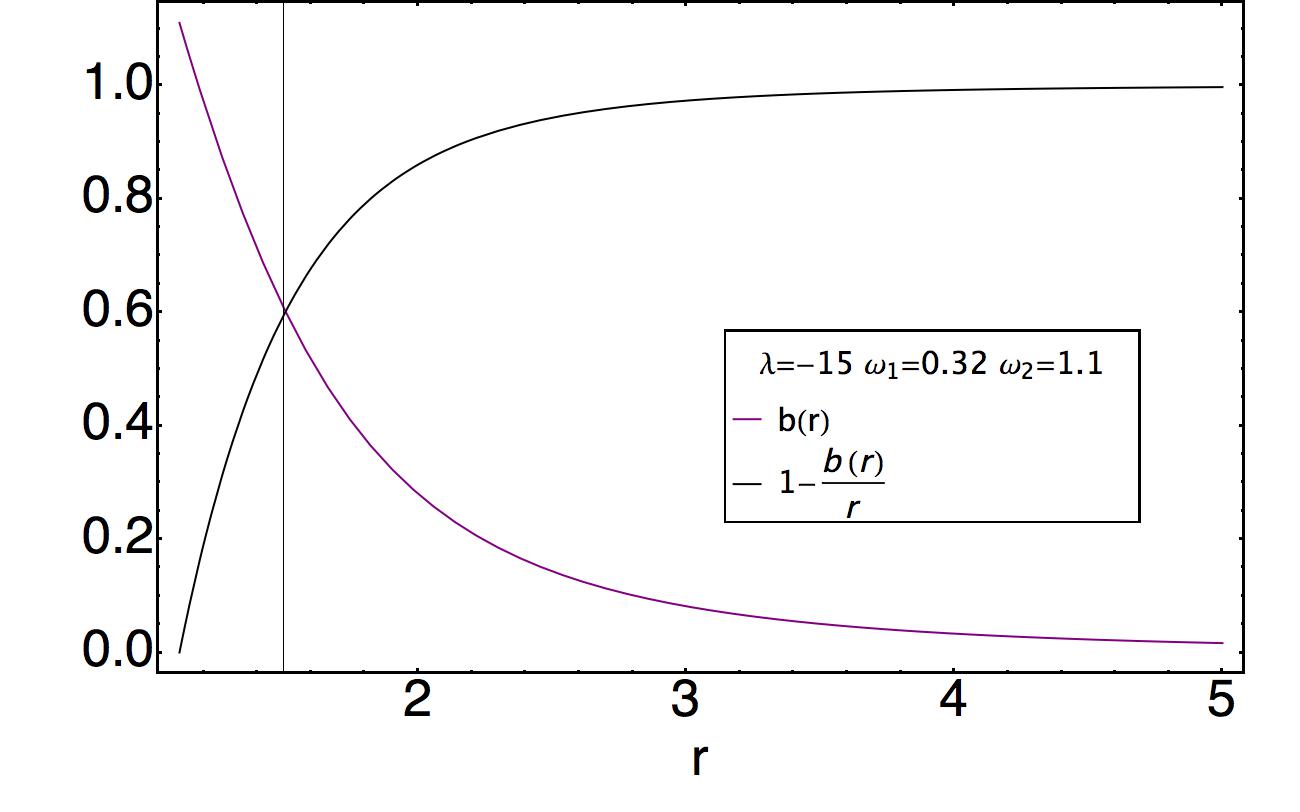} &&
\includegraphics[width=80 mm]{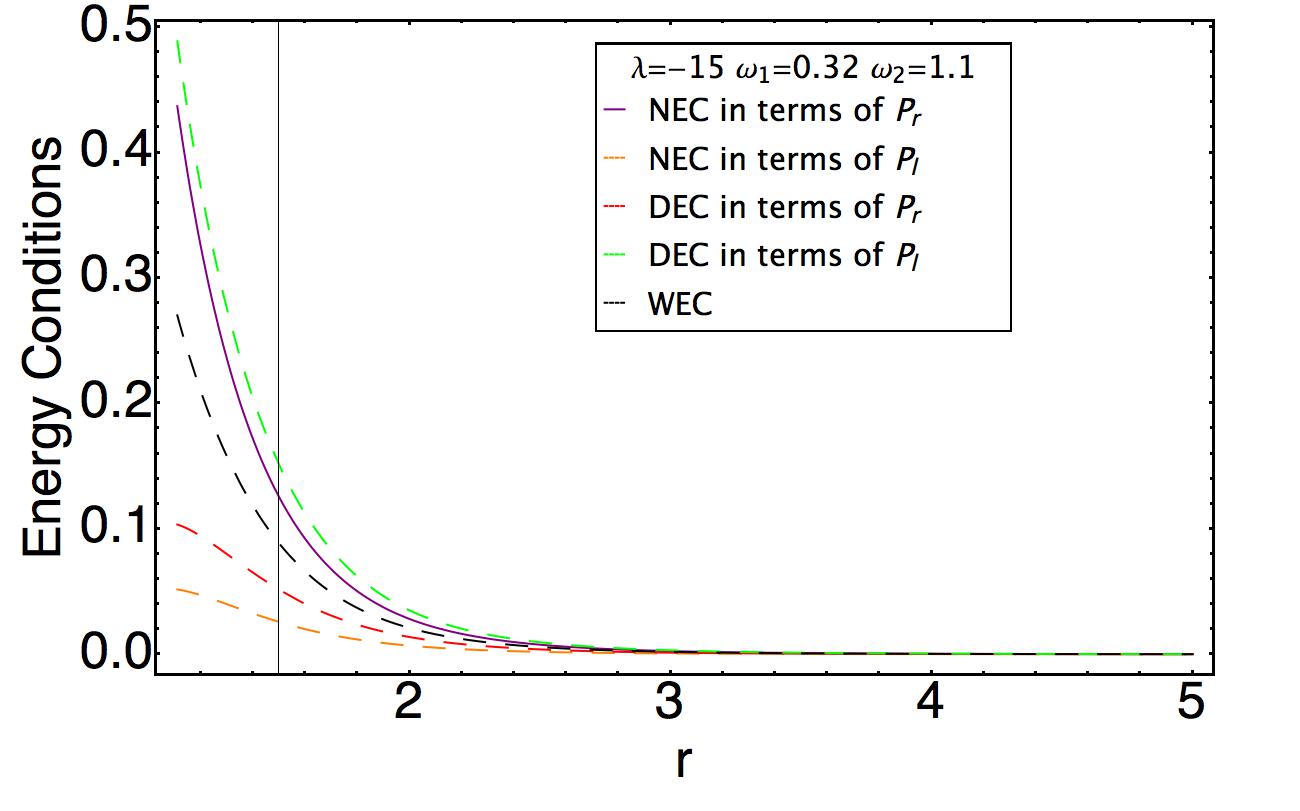}
 \end{array}$
 \end{center}
\caption{Graphical behavior of the shape function $b(r)$ for model $3$ (left plot). The same plot demonstrates that the solution for $b(r)$ satisfies $1- b(r)/r > 0$, for $r > r_{0}$. The right plot shows the validity of the WEC, DEC, and NEC for $ \lambda = -15,~\omega_{1} = 0.32$ and $\omega=1.1$. $r$ is in $[km]$.}
 \label{fig:Fig5}
\end{figure}

One interesting wormhole solution, with this particular type of matter, has been obtained for $\lambda =-15,$ $\omega_{1}=0.32$ and $\omega_{2}=1.1$. Moreover, in this case the throat of the wormhole occurs at $r_{0}=1.11$. The graphical behavior of the shape function $b(r)$ is depicted on the left plot of Fig.~\ref{fig:Fig5}. It should be remarked that we have obtained a solution satisfying  all the conditions mentioned in section~\ref{sec:WMFE}. Study of the energy conditions as given on the rhs plot of  Fig.~\ref{fig:Fig5} proves their validity. On the other hand, an extended analysis concludes that the values of $\omega_{1}$ and $\omega_{2}$ can just violate the DEC in terms of $P_{l}$ and $P_{r}$. Anyhow, in these cases the NEC and WEC will still be valid, with  $\rho + P_{r} + 2P_{l} = 0$.

\begin{figure}[h!]
 \begin{center}$
 \begin{array}{cccc}
\includegraphics[width=80 mm]{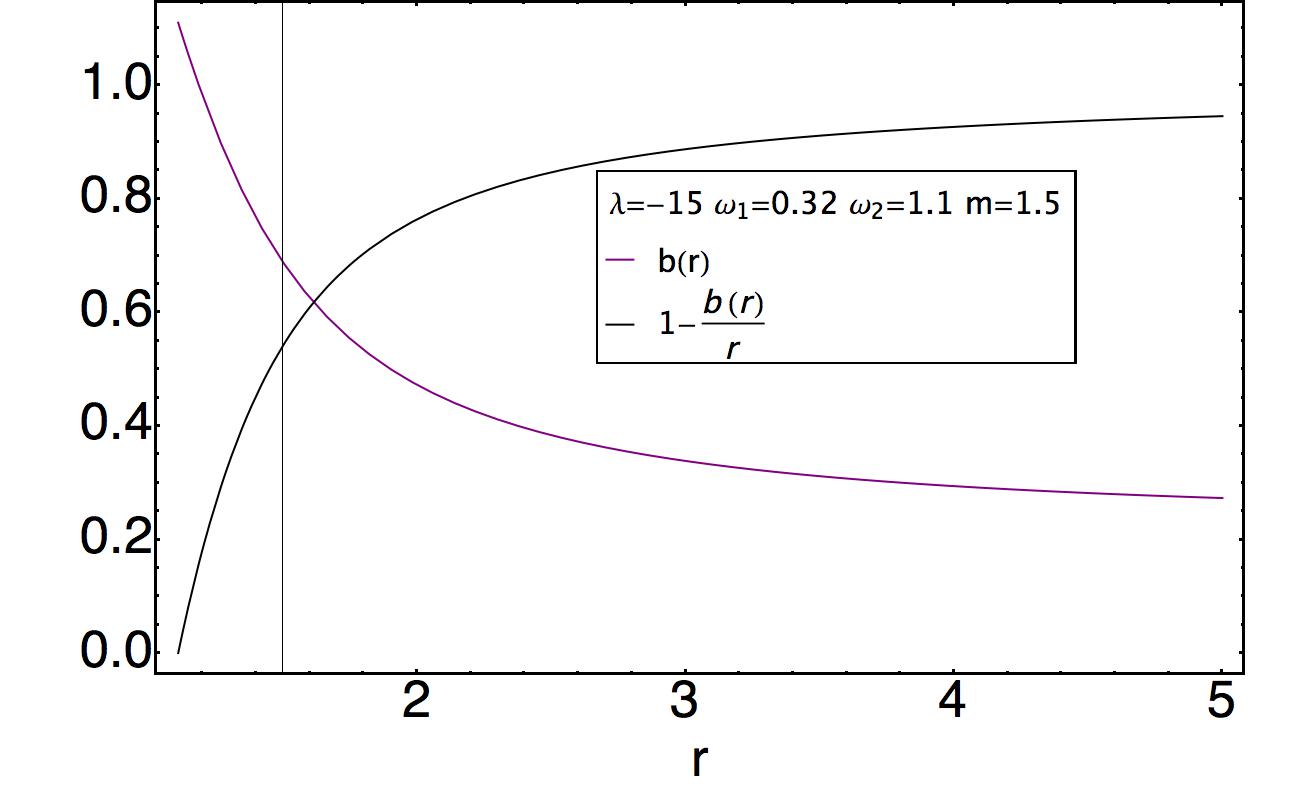} &&
\includegraphics[width=80 mm]{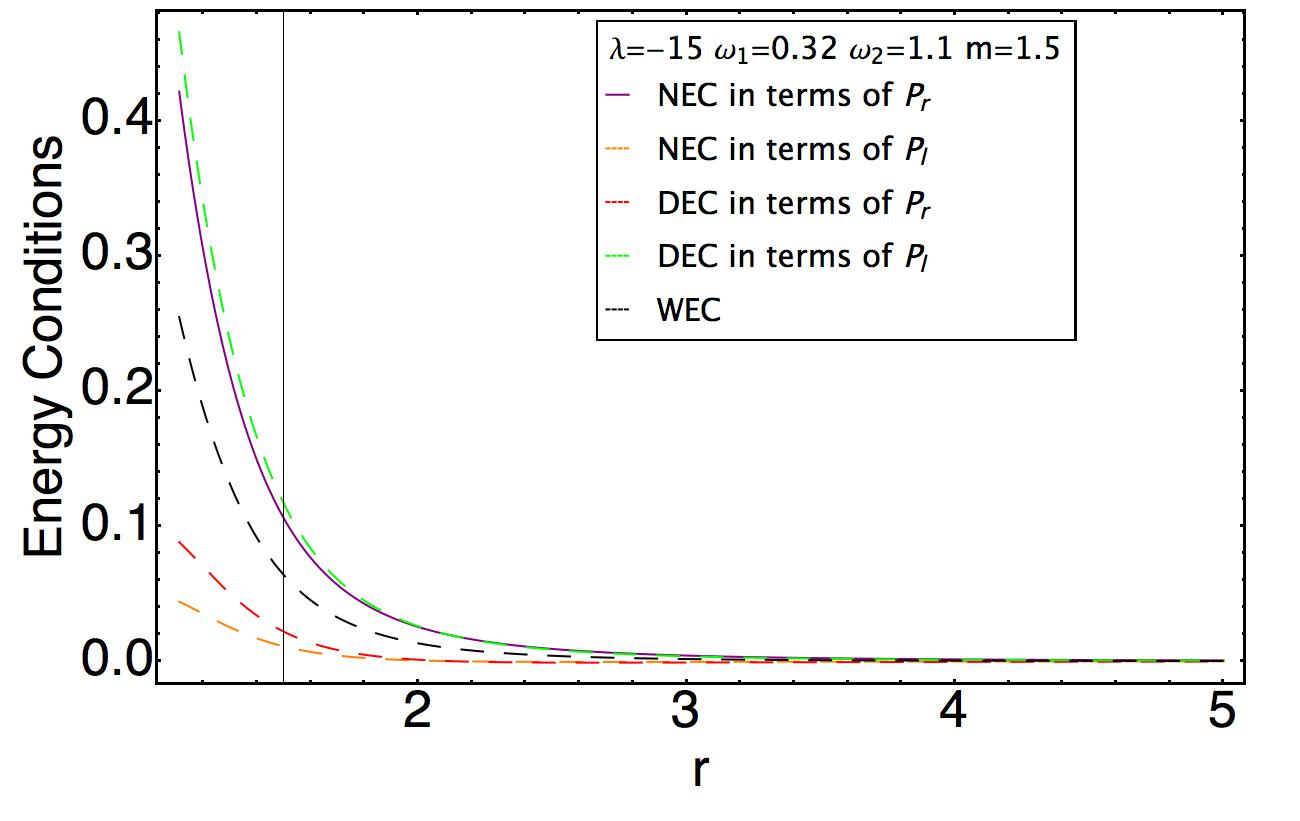}\\
\includegraphics[width=80 mm]{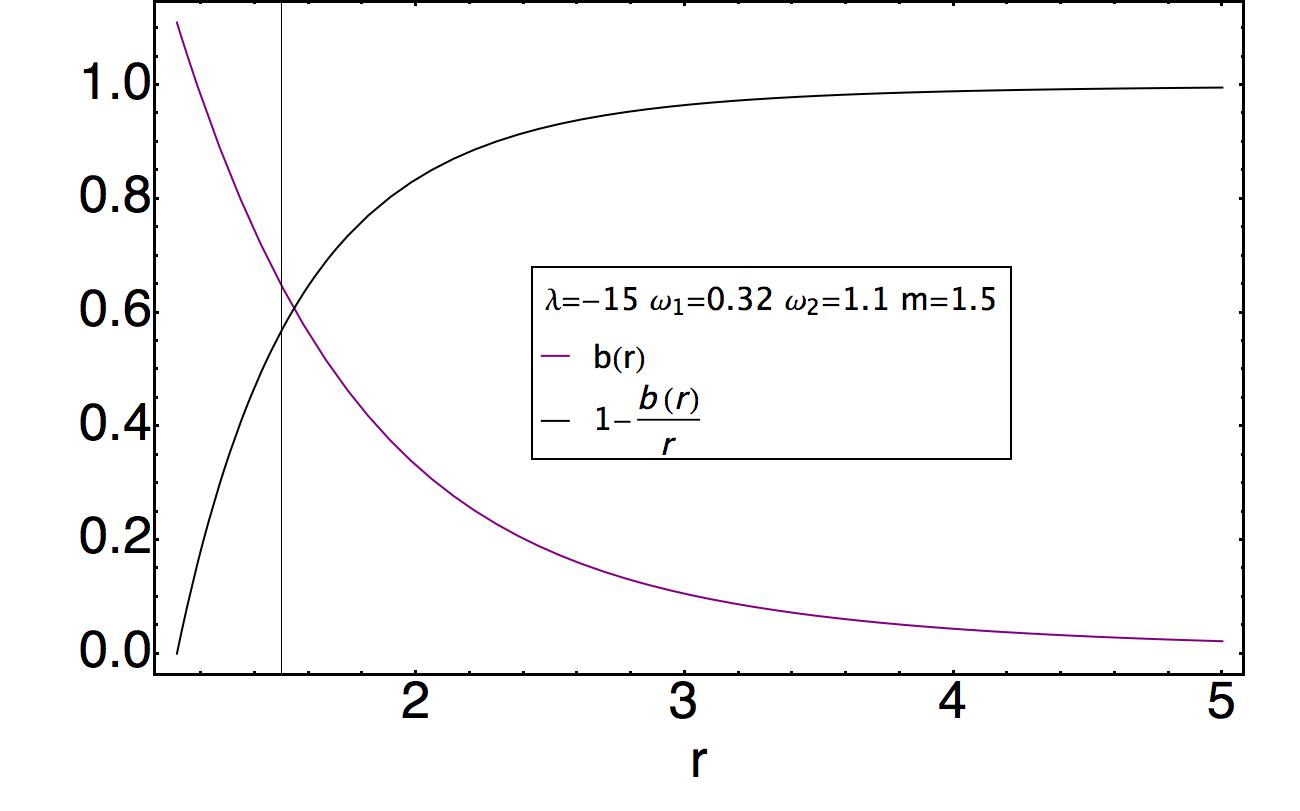} &&
\includegraphics[width=80 mm]{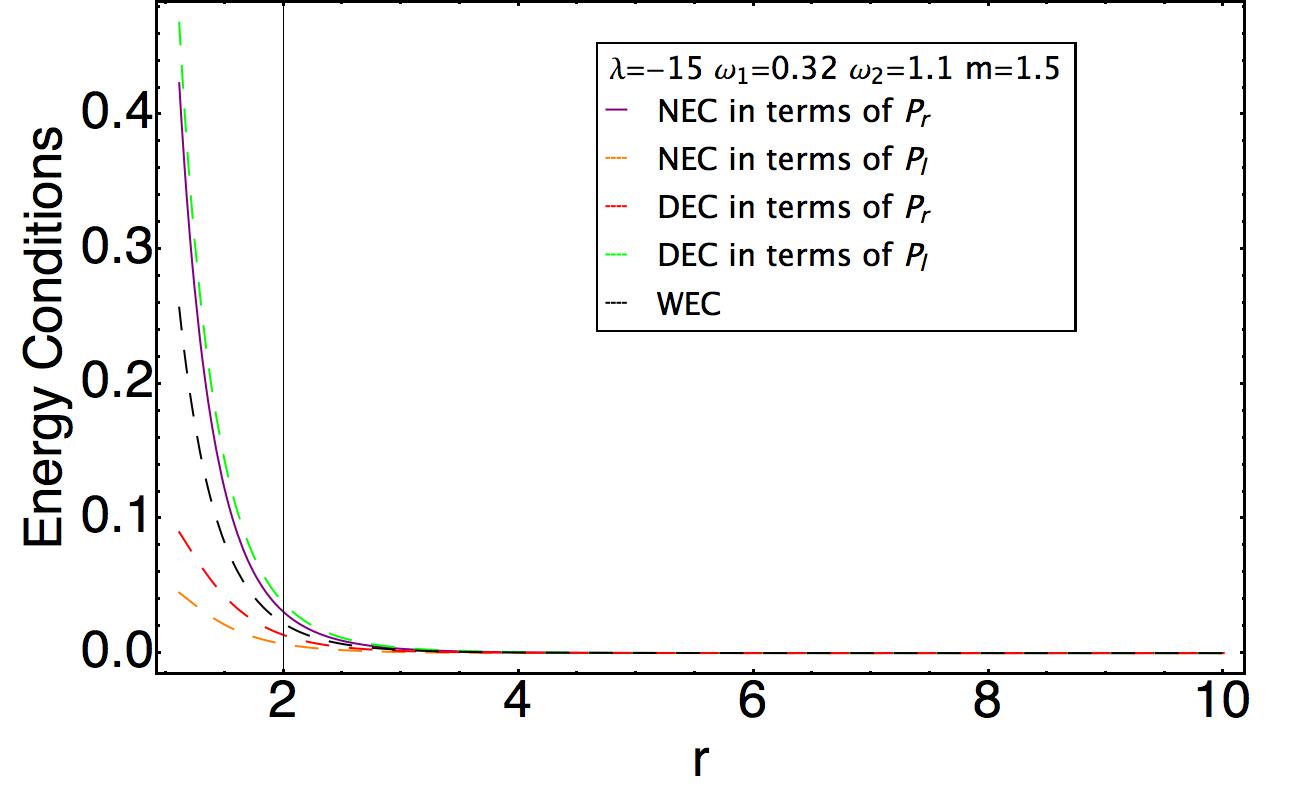}
 \end{array}$
 \end{center}
\caption{Graphical behavior of the shape function $b(r)$ for the model described by Eq.~(\ref{eq:Fluid1}) (upper-left plot). From the plot we see that the solution for $b(r)$ satisfies $1- b(r)/r > 0$, for $r > r_{0}$. The upper-right plot shows the validity of the WEC, DEC, and NEC for $ \lambda = -15,~\omega_{1} = 0.32,~\omega=1.1$ and $m=1.5$. The bottom plot depicts the corresponding behavior, for the alternative model given by Eq.~(\ref{eq:Fluid2}). $r$ is in $[km]$, while pressure and the energy densities are dimensionless.}
 \label{fig:Fig6}
\end{figure}

Using numerical analysis, we have also obtained wormhole solutions for the following two equations of state describing the matter of the wormhole:
\begin{equation}\label{eq:Fluid1}
P_{r} = \omega_{1} r^{m} \rho + \omega_{2} \rho^{2},
\end{equation}
and
\begin{equation}\label{eq:Fluid2}
P_{r} = \omega_{1} \rho + \omega_{2}r^{m} \rho^{2}.
\end{equation}
The graphical behavior of $b(r)$ and the energy conditions for both cases are given in Fig.~\ref{fig:Fig6}. For both cases, the corresponding solutions had been obtained for $\lambda =-15, \omega_{1}=0.32, \omega_{2}=1.1, m = 1.5$, and the wormhole develops at $r_{0}=1.11$. These numerical solutions are interesting since they provide valid energy conditions. Moreover, during the numerical analysis we have seen that the parameter space for all three cases can be divided into several regions, where some of the energy conditions remain valid and some of them are violated. From this perspective, since the EoS of the wormhole matter  is not very well constrained and understood, we are not able to definitely indicate which of the observed scenarios corresponds to a real, feasible case. On the other hand, we have obtained new wormhole models with a very rich spectrum of possible behaviors that according to the new observational data can be constrained again, with the final aim of assessing their viability. 

\section{Wormhole models with $R+\alpha R^{2} + 2 f(T)$ gravity}\label{sec:R2} 

In the preceding sections, we  have already mentioned  some good reasons why is it interesting to consider theories of gravity with a modified matter part. In fact, in the recent literature, there is a good amount of excellent studies justifying the consideration of different forms of $f(R)$ gravity~(see appropriate references at the end of this paper). The need to take into account quantum effects can also be invoked as a reason to consider modifications of the geometrical part of the gravity equations. The most important example in this direction is the very well know $R+\alpha R^{2}$ Starobinsky model~\cite{Starobinsky:1980}. This is the model of inflation preferred by theoreticians. The wormhole solutions considered in Sects.~\ref{sec:MEW} and~\ref{seq:M2} are based on a very simple modification of this model, which actually simplifies the resulting field equations. In any case, even with these simplified field equations, one realizes that to construct exact wormhole models for a chosen EoS can still be very difficult. This is the main reason why we will have to deal numerically with the wormhole models under  study there,  and only a few of them will be obtained in an analytical way. 

The goal in this section is to study wormhole models corresponding to the theory
\begin{equation}
f(R,T) = R + \alpha R^{2} + 2 f(T),
\end{equation}
where the same assumptions  taken into account earlier in this paper are implied.  Before starting the analysis in this case, let us derive the expressions of some key quantities obtained from the definitions for the wormhole metric Eq.~(\ref{eq:WHMetric}), which we will use. In particular, the form of the Ricci scalar in this case reads 
\begin{equation}
R = \frac{2 b^{\prime}}{r^{2}}.
\end{equation} 
On the other hand, for $\Box f_{R}$, we have
\begin{equation}
\Box f_{R} = \left ( 1-\frac{b}{r} \right ) \left( \frac{f^{\prime}_{R}}{r} + f^{\prime \prime}_{R}  + \frac{f^{\prime}_{R} (b - r b^{\prime})}{2r^{2} (1 - b/r)} \right),
\end{equation} 
while
\begin{equation}
\nabla_{1}\nabla_{1} f_{R} = \frac{f^{\prime}_{R}(b - r b^{\prime})}{2r^{2} (1-b/r)} + f^{\prime \prime}_{R}, 
\end{equation}
\begin{equation}
\nabla_{2} \nabla_{2}  f_{R}  = r\left ( 1-\frac{b}{r} \right ) f^{\prime}_{R},
\end{equation}
$\nabla_{0} \nabla_{0}  f_{R}  = 0$ and $\nabla_{3} \nabla_{3}  f_{R}  = r\left ( 1-\frac{b}{r} \right ) f^{\prime}_{R} \sin^{2}\theta$. After some algebra, we obtain
\begin{equation}
\frac{b^{\prime}}{r^{2}} = 8\pi \rho - \frac{\alpha}{2}R^{2} - \lambda T + \Box f_{R},
\end{equation}   
$$
-\frac{b}{r^{3}} = 8 \pi P_{r} + 2 \lambda (P_{r} + \rho) + \frac{\alpha}{2}R^{2} + \lambda T + 2 \alpha R \left( \frac{b-rb^{\prime}}{r^{3}}\right) + $$
\begin{equation}
+ \frac{b - rb^{\prime}}{2r^{2}} f^{\prime}_{R} + \left( 1 - \frac{b}{r}\right) f^{\prime \prime}_{R} - \Box f_{R},
\end{equation}
and
\begin{equation}
\frac{b - r b^{\prime}}{2r^{3}} = 8 \pi P_{l} + 2 \lambda(P_{l} + \rho) - \alpha R \frac{b+r b^{\prime}}{r^{3}} + \frac{\alpha}{2}R^{2} + \lambda T + \frac{1}{r} \left( 1 - \frac{b}{r}\right) f^{\prime}_{R} - \Box f_{R}.
\end{equation}

\begin{figure}[h!]
 \begin{center}$
 \begin{array}{cccc}
\includegraphics[width=80 mm]{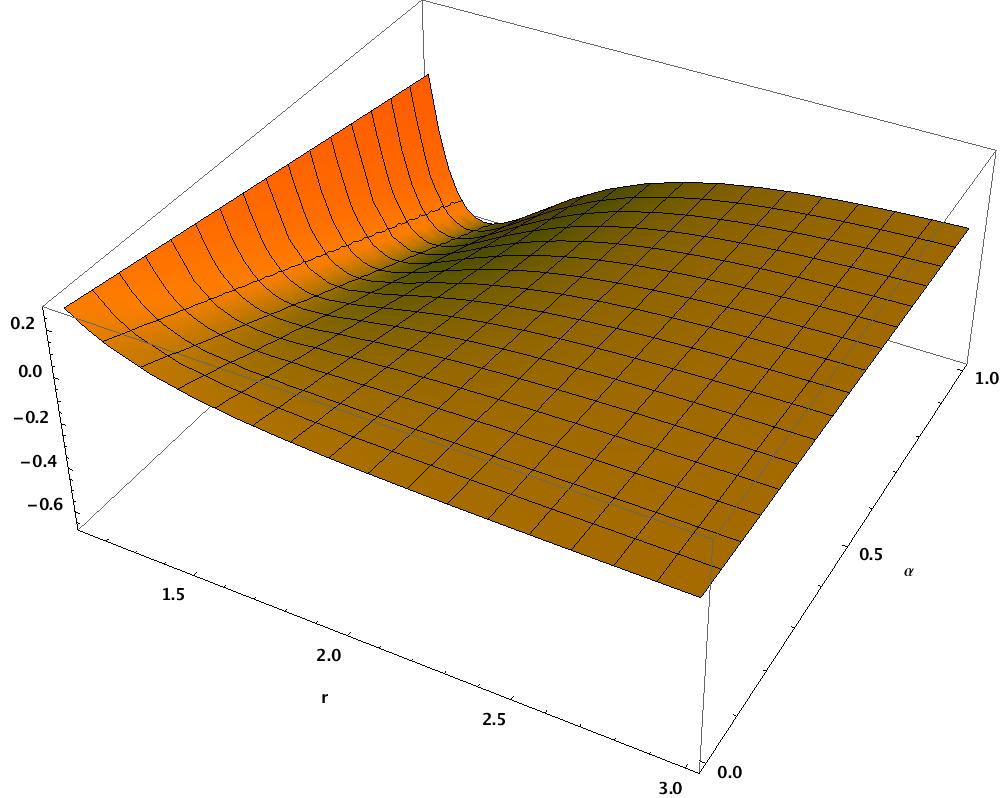}
 \end{array}$
 \end{center}
\caption{Graphical behavior of the null energy condition~(NEC i.e $\rho + P_{r}$) in terms of the pressure  $P_{r}$ for the model with $P_{r} = \omega_{1} \rho + \omega_{2} \rho^{2}$, for different values of $\alpha$. $r$ is in $[km]$, while pressure and the energy densities are dimensionless.}
 \label{fig:Fig7}
\end{figure}

From the three equations above it is easy to see that even if we will assume a relation, for instance, between $P_{l}$ and $P_{r}$ as it has been done already, then in order to find $b(r)$ we need to solve a third-order differential equation. In general, for some simple cases we can expect to find exact solutions. In addition to this, imposing the assumptions concerning the EoS leaves us, generically, only one option, namely to study the models numerically. However, in the case of the considered form for the shape function $b(r)$, it is  possible to obtain exact expressions for $\rho$, and the $P_{r}$ and $P_{l}$ pressures. For instance, in the case when $b = r_{0}/r$ with EoS given by Eq.~(\ref{eq:Fluid0}), we can obtain a wormhole solution~(two solutions, actually), with
\begin{equation}\label{eq:rhoFloud0_R2}
\rho_{1,2} = -\frac{  \omega_{1} + 1 \mp  \frac{ A_{0} }{(\lambda +4 \pi ) r^{8} } }{2 \omega_{2} },
\end{equation}
and 
\begin{equation}\label{eq:PlFloud0_R2}
P^{1,2}_{l} = \frac{B_{0}}{2\lambda} \pm \frac{A_{0} } {2 \omega_{2} \lambda r^8},
\end{equation} 
where $A_{0} = \sqrt{(\lambda +4 \pi ) r^8 \left((\lambda +4 \pi ) r^8 (\omega_{1}+1)^2-4 r_{0} \omega_{2} \left(r^4-40 \alpha  r^2+44 \alpha  r_{0}\right)\right)}$, $B_{0} = -\frac{(\lambda +4 \pi ) (\omega_{1}+1)}{\omega_{2}}+\frac{2 \alpha  (61 \lambda +156 \pi ) r_{0}^2}{(\lambda +4 \pi ) r^8}-\frac{8 \alpha  (13 \lambda +32 \pi ) r_{0}}{(\lambda +4 \pi ) r^6}+\frac{2 (\lambda +2 \pi ) r_{0}}{(\lambda +4 \pi ) r^4}$. From the above equations it is easy to see that, for instance, for $\omega_{2} < 0$ and $\omega_{1} > - 1 \pm  \frac{ A_{0} }{(\lambda +4 \pi ) r^{8} } $, we can take $\rho_{1,2} > 0$.

Fig.~\ref{fig:Fig7}  shows the behavior of $\rho + P_{r}$ obtained  from Eqs.~(\ref{eq:rhoFloud0_R2}) and~(\ref{eq:PlFloud0_R2}), for $\rho_{1}$ and $P^{1}_{l}$. In particular, we have observed, that the parameter  $\alpha$ in front of $R^{2}$ can actually affect the null energy condition. The plot in Fig.~\ref{fig:Fig7}  shows explicitly that, for some values of the parameter, the NEC in terms of $P_{r}$ can be violated, for some values of $r$. However, this is a local violation, which induces also local violations of the DEC and WEC in terms of both pressures. This is a very interesting situation, since we have seen above that in the case $f(R) = R$ we can have a wormhole solution satisfying all energy conditions. The values of the parameters are fixed exactly as it has been done for the same fluid model when dealing with the case  $f(R) = R$.   

\subsection{Models with $\omega_{i} \propto r^{m}$}

Now, let us briefly discuss the results obtained for the other two fluid models. The way we consider the  shape functions allows us obtain expressions for $P_{r}$, $P_{l}$, and $\rho$. In particular, for the model given by Eq.~(\ref{eq:Fluid1}), we obtain
\begin{equation}\label{eq:rhoFluid2R2}
\rho_{1,2} = -\frac{\omega_{1} r^{m}+1 \pm \frac{A_{1}} {(4\pi + \lambda)r^{8}}}{2 \omega_{2}}
\end{equation}     
and
\begin{equation}\label{eq:PlFluid2R2}
P^{1,2}_{l} =  \frac{B_{1}}{2\lambda} \mp \frac{A_{1}}{2\lambda \omega_{2} r^{8}},
\end{equation} 
where $A_{1} = \sqrt{(\lambda +4 \pi ) r^8 \left((\lambda +4 \pi ) r^8 \left(\omega_{1} r^m+1\right)^2-4 r_{0} \omega_{2} \left(r^4-40 \alpha  r^2+44 \alpha r_{0}\right)\right)}$, $B_{1} = -\frac{(\lambda +4 \pi ) \left(\omega_{1} r^m+1\right)}{\omega_{2}}+\frac{2 \alpha  (61 \lambda +156 \pi ) r_{0}^2}{(\lambda +4 \pi ) r^8}-\frac{8 \alpha  (13 \lambda +32 \pi ) r_{0}}{(\lambda +4 \pi ) r^6}+\frac{2 (\lambda +2 \pi ) r_{0}}{(\lambda +4 \pi ) r^4}$.

A careful study leads to the conclusion that, for the  case of the wormhole solution given by $\rho_{1}$ and $P^{1}_{l}$ of Eqs.~(\ref{eq:rhoFluid2R2}) and~(\ref{eq:PlFluid2R2}), all energy conditions will be  locally violated, and that this violation occurs far from the throat of the wormhole. Again, as in the case of the previous model, this violation comes from the $\alpha$ parameter of the $R^{2}$ term. On the other hand, if we consider the wormhole solution given by $\rho_{2}$ and $P^{2}_{l}$ of Eqs.~(\ref{eq:rhoFluid2R2}) and~(\ref{eq:PlFluid2R2}), respectively, then even for the values of $\alpha$ for which the NEC in terms of $P_{r}$ is valid, other energy conditions are violated. A similar situation has been observed also for the model given by Eq.~(\ref{eq:Fluid2}), for which we have obtained
\begin{equation}
\rho_{1,2} = \frac{r^{-m-8} \left(  -(\lambda +4 \pi ) r^8 (\omega_{1}+1)\right) \pm A_{2} } {2 (\lambda +4 \pi ) \omega_{2}},
\end{equation}
and
\begin{equation}
P^{1,2}_{l} = \frac{B_{2}}{2\lambda} \pm \frac{r^{-8-m} A_{2}}{2\lambda \omega_{2}} .
\end{equation}
$A_{2} = \sqrt{(\lambda +4 \pi ) r^8 \left((\lambda +4 \pi ) r^8 (\omega_{1}+1)^2-4 r_{0} \omega_{2} r^m \left(r^4-40 \alpha  r^2+44 \alpha  r_{0}\right)\right)}$ and $B_{2} = \frac{2 \lambda  r_{0} \left(r^4-52 \alpha  r^2+61 \alpha r_{0}\right)+4 \pi  r_{0} \left(r^4-64 \alpha  r^2+78 \alpha r_{0}\right)}{(\lambda +4 \pi ) r^8}-\frac{(\lambda +4 \pi ) (\omega_{1}+1) r^{-m}}{\omega_{2}}$.

\section{Discussion and Conclusions}\label{sec:Discussion}

In this paper we have constructed a number of wormhole models corresponding to $f(\textit{R}, \textit{T})$ extended theories of gravity, with $f(\textit{R}, \textit{T}) = R + \lambda T$, being $T = \rho + P_{r} + 2P_{l}$ the trace of the energy momentum tensor. This particular modification of gravity comes from extra  contributions to the matter part. They could be  related to the existence of imperfect fluids. On the other hand, quantum effects, such as those involved in particle production, can be another important motivation to consider modified theories of gravity with different  matter content. 

Two exact wormhole models have been constructed, assuming  the following relations between the radial and the lateral pressures $P_{l} = n P_{r} +  \alpha P_{r}^2$ and $P_{l} = n P_{r} +  \alpha r^{m} P_{r}^2$. For  the first model, in particular, we have seen that the NEC in terms of the radial and DEC in terms of the lateral pressures can be violated, while the WEC in terms of $P_{l}$ is still fulfilled. On the other hand, in the case of the model with $P_{l} = n P_{r} +  \alpha r^{m} P_{r}^2$, we have seen that only the  NEC in terms of the radial pressure can be violated. 

In the second part of the paper we have numerically constructed, in addition, three other wormhole models by assuming that  $P_{r} = \omega_{1} \rho + \omega_{2} \rho^{2}$, where $\omega_{1}$ and $\omega_{2}$ can be either constant or depend on $r$. In the case of non-constant $\omega_{i}$ models, we have focussed our attention on the following form for $\omega_{i} \propto r^{m}$. Detailed numerical study of these three models led to the conclusion  that we can have wormhole solutions, and that it is possible to satisfy all energy conditions. Moreover, the parameter space can be divided into several regions, where some of the energy conditions are still valid. For instance, we have observed that, for appropriate values of the parameters, we can obtain a wormhole solution with valid NEC, WEC, and SEC, while only DEC is violated. On the other hand, we observed that for negative $\omega_{1}$ and $\omega_{2}$ it is possible to obtain wormhole solutions satisfying all energy conditions at the throat of the wormhole, albeit they are bound to be violated, locally, far from the throat. 

Finally, we have also considered wormhole solutions constructed numerically for the fluids given by $P_{r} = \omega_{1} \rho + \omega_{2} \rho^{2}$, where $\omega_{1}$ and $\omega_{2}$ can be either constant or depend on $r$~($\omega_{i} \propto r^{m})$ in the case of $f(R) = R + \alpha R^{2}$. For a specific form of the shape function, namely  $b(r) = r_{0}/r$, we observed that one of the solutions describes a wormhole for which only the NEC in terms of $P_{r}$ is violated, namely locally and far from the throat. On the other hand, other energy conditions are violated everywhere for appropriate values of the parameters of the  model. However, for the other solution we have shown, that if the parameters of the model are such that all energy conditions are valid, in the case of $f(R) = R$, then, now in the case of $f(R) = R + \alpha R^{2}$ the energy conditions are locally violated, owing to the parameter $\alpha$.  

We should note that all our solutions are functions of the $\lambda$ parameter  appearing in the corresponding $F(R,T)$ theory considered, and also on the
parameters defining the fluids, so that they will be generic functions of the new theories ($T$ plays a definite role). Also, It can be seen from the shape functions for each case that we obtained results, which are different from those of the $F(R)$ case, being the specific cases discussed in correspondence with  particular values of $\lambda$. In addition, the traversability of
the wormholes strongly depends on those values of $\lambda$, and this cannot be reached by simply adjusting the values of the parameters defining the fluids.

Now, let us summarize some perspectives concerning possible future studies in order to demonstrate the validity of the  wormhole solutions here obtained. The present work is in principle a purely theoretical one, demonstrating the possibilities of new departures from previous studies existing in the recent literature. The resulting solutions are interesting in view of future development to be pursued in this direction and closely related to the direct or indirect observational detection of wormholes. The confirmation of the wormhole detection is not only important for constraining the wormhole matter EoS, but  it could be also used in order to constrain underlying theories of gravity. The new data from the cosmological observations available today will provide very tight constraints on the underlying theories of gravity. Moreover, the confirmation of the wormhole detection can bring new results into particle physics and physics of gravitational waves. 

One of the possibility to detect wormholes could be related to its lensing properties, as compared to a huge amount of high-quality observational data existing today. The lensing data has been intensively used in the recent literature with the aim to constrain different theories of gravity and related cosmological models. From this perspective, we have a real possibility to impose tight constraints on the wormhole matter equation and on the wormhole physics. Also, it should be mentioned that another interesting approach in this direction would be the study of the particle creation in wormhole space-times. The existing limitations with this approach, again, are related to the undetected nature of the wormholes. Theoretically, we can use lensing data and study the particle creation which, in our opinion it can affect the lensing properties of wormholes at least in two ways. In particular, during particle creation, the wormhole can become unstable in one case due to the direct process of the particle creation and in the second case due to feedback from the metric, owing to particle creation. Moreover, more complicated and interesting situation could be observed steming from the highly non-linear physics, establishing some interplay between these two cases. One can assume also that the physics developed in such scenarios would be observed on  cosmological scales, for instance in terms of a negative pressure i.e. as a contribution to dark energy.  We believe that in all mentioned cases the lensing properties of the wormhole will be changed and the observational data would be able to provide valuable hints on this issue. Moreover, in this scenario, another key aspect is correctly to closely model the particle creation rate. 

Another important issue one must address in wormhole theories is that of the possible nonconservation of the matter energy-momentum tensor. 
It is known that in case of modified theories of gravity the validity of the energy conditions can be achieved due to the extra term appearing, related to the modification of gravity. However, the matter contribution itself still can violate the energy conditions. In this regard, the extra term coming from the modification of gravity may act as a source and may be used to mediate particle creation. Eventually, in order to study this particle creation, one could resource to the generalized Tolman-Oppenheimer- Volkoff equation, which provides the conditions under which the wormhole solution is stable. Moreover, the balance violation between the gravitational, hydrostatic, and anisotropic forces would be a direct hint towards a rigorous formulation of wormhole physics. We are aware of these situations, which lie beyond the scope of the present work and will be addressed in future work.

We must also mention that our initial analysis shows that the wormhole solutions here  obtained can indeed be stable, due to the balance that can be stablished between the gravitational, hydrostatic, and anisotropic forces. Further detailed studies in this direction, including the above considered one of particle creation in wormhole space-time metrics, for the cases here studied and their effects on the lensing properties of the wormholes will be the subject of a separate research. At this stage of the study, since the EoS is not well understood and constrained, we cannot definitely conclude which one of the solutions presented in this paper is the most feasible as a working model for the cosmos. We hope that in the near future, with the help of more observational data, including lensing data, of substantially better quality, we will be able to assess the final validity of the models considered in this paper. Moreover, in forthcoming works, we expect to report on further analysis of these fluids, extended to include other forms of the shape function in order to better constrain the viability domain of the wormhole solutions.

\section*{Acknowledgements}
We are grateful for the detailed comments of an anonymous referee, which led to a definite improvement of the first version of this manuscript. 
EE has been supported in part by MINECO (Spain), Project FIS2016-76363-P,  by the CPAN Consolider Ingenio 2010 Project, and by the Catalan Government, Project AGAUR 2017-SGR-247.
MK is supported in part by the Chinese Academy of Sciences President's International Fellowship Initiative Grant (No. 2018PM0054).

\end{document}